# Defect migration in supercrystalline nanocomposites

Dmitry Lapkin[1,*,i], Cong Yan[2], Emre Gürsoy[3], Hadas Sternlicht[4], Alexander Plunkett[5], Büsra Bor[5], Young Yong Kim[1], Dameli Assalauova[1,iii], Fabian Westermeier[1], Michael Sprung[1], Tobias Krekeler[6], Surya Snata Rout[6,ii], Martin Ritter[6], Satishkumar Kulkarni[7], Thomas F. Keller[7,8], Gerold A. Schneider[5], Gregor B. Vonbun-Feldbauer[3,9], Robert H. Meissner[3,9], Andreas Stierle[7,8], Ivan A. Vartanyants[1], Diletta Giuntini[2,5,*]

[1] Photon Science, Deutsches Elektronen-Synchrotron (DESY), Hamburg, Germany

[2] Department of Mechanical Engineering, Eindhoven University of Technology, Eindhoven, Netherlands

[3] Institute of Interface Physics and Engineering, Hamburg University of Technology, Hamburg, Germany

[4] Department of Materials Science and Engineering, Pennsylvania State University, University Park, PA, USA

[5] Institute of Advanced Ceramics, Hamburg University of Technology, Hamburg, Germany

[6] Electron Microscopy Unit, Hamburg University of Technology, Hamburg, Germany

[7] Centre for X-ray and Nano Science, Deutsches Elektronen-Synchrotron (DESY), Hamburg, Germany

[8] Department of Physics, University of Hamburg, Hamburg, Germany

[9] Institute of Surface Science, Helmholtz-Zentrum-Hereon, Geesthacht, Germany

Supercrystalline nanocomposites (SCNCs) are nanostructured hybrid materials with unique emergent functional properties. Given their periodically arranged building blocks, they also offer interesting parallelisms with crystalline materials. They can be processed in multiple forms and at different scales, and crosslinking their organic ligands via heat treatment leads to a remarkable boost of their mechanical properties. This study shows, via X-ray and in-situ scanning transmission (STEM) electron microscopy analyses, how each of these processing steps plays a distinct role in the generation, migration, interaction and healing of supercrystalline defects. Pressing of SCNCs into bulk pellets leads to a distortion of the otherwise *fcc* superlattice, while emulsion-templated self-assembly yields supraparticles (SPs) with stacking faults and size-dependent symmetries. Interestingly, heat treatment at the same temperatures as those applied for the organic crosslinking has significant effects on planar defects. Stacking faults migrate and get healed, as also confirmed via molecular dynamics simulations, and inter-supercrystalline "grain" boundaries undergo structural changes. These rearrangements of defects at the supercrystalline scale (tens of nm) in nanocomposites with such remarkable mechanical properties (compressive strength of 100-500 MPa) provide new insights into the formation and evolution of ordered assemblies of functionalized nanoparticles.

[*] Corresponding authors

[i] Current address: Institute of Applied Physics, University of Tübingen, Auf der Morgenstelle 10, 72076 Tübingen, Germany

[ii] Current address: School of Earth & Planetary Sciences, National Institute of Science Education and Research, Jatani, 752050 Khurda, India; Homi Bhabha National Institute, Training School Complex, Anushaktinagar, 400094 Mumbai, India

[iii] Current address: Constructor University, Campus Ring 1, D-28759 Bremen, Germany

**Introduction**

Supercrystalline nanocomposites (SCNCs) are new remarkable materials, consisting of self-assembled inorganic nanoparticles (NPs) that are surface-functionalized with organic ligands.[1–6] This combination of nano-sized building blocks and their long-range order arrangement, analogous to that of atoms in crystals, is emerging as a powerful material design strategy. By tailoring composition, NP size, arrangement and spacings, emergent collective properties can be fostered, with promising applications in e.g. the catalysis, energy, optoelectronics and magnetic materials fields.[1,4,7–10]

Beyond this broad spectrum of potential applications, SCNCs are also studied in light of available knowledge on conventional crystalline materials, with which parallelisms are being drawn.[11] Crystalline structures, phases and defects have long been investigated with systems of periodically arranged building blocks at larger scales – starting with bubble rafts,[12] through colloidal crystals of unfunctionalized microparticles,[13] all the way to the recent progress made with bimodal distributions of DNA-functionalized NPs.[14] By fine-tuning NP interactions and architectures, new paths towards high-tech programmable materials are being paved.

Two main issues, however, still stand in the way of SCNCs' implementation into devices: controlling their assembly into scales beyond 2D materials and micro-sized domains, and increasing their mechanical robustness.[5,15] On the former aspect, progress is being made by self-assembly controlled via targeted ligand interactions;[16] conducting self-assembly in macro-scale geometries followed by a pressing processing step;[17,18] or via hierarchical designs, which employ SCNCs as micro-building blocks for macroscopic materials.[19,20] On the mechanical properties side, thermally-induced organic crosslinking has been proven to be a very effective mean towards strengthening, stiffening, hardening and even toughening SCNCs.[5,17,21–27]

As in any material system, all these aspects are strongly influenced by defects. Defects control mechanical and functional performance of a material, and they are thus another feature to tune in the development of programmable materials.[28] Research on how processing affects defects in SCNCs, and how defects in turn affect material properties and performance, is however still in its infancy. Most studies have focused on the occurrence of imperfections during the assembly of colloidal crystals of unfunctionalized microparticles.[29–38] A few have considered mechanically-induced defects (and thus superlattice imperfections that can be induced during service instead of during processing).[39] The important aspects of defect mobility, migration, interaction and healing have been addressed so far mainly for single-component colloidal crystals, i.e. periodic arrays of unfunctionalized microparticles.[40] Most of these studies focus on defect migration during self-assembly,[40] and only a few on the post-assembly stages, by looking into "melting" or "viscous flow", and in the effects of dopants, gravitational effects, or optical tweezers.[41–49] In all of these cases, the colloidal crystals are in a soft matter state, which is significantly different with respect to the strong SCNCs.[5]

In the case of SCNCs, most investigations on superlattice defects focus on their formation during self-assembly in thin films[50,51] or micro-sized single supercrystals.[52,53] Mechanically induced defects are just starting to be analysed, while their migration remains unexplored.[24,54] Annealing (heat-treating) SCNCs is, on the other hand, starting to provide meaningful insights on the occurrence of strain or phase transitions in superlattices.[55,56] So far these transitions have been addressed via superlattice "melting" in temperature ranges well below 100 ºC, and mainly for DNA-based NP superlattices.[57–60]

SCNCs, however, are processable in a broad spectrum of compositions and in states that go well beyond soft matter. Once the organic ligands are crosslinked, the high values of strength, hardness and elastic modulus that they reach fully qualify these materials as hard composites, withstanding temperatures up to 350 ºC without undergoing degradation.[5,21,25] This kind of ultra-strong SCNC is processed in two- or three-step routines, involving self-assembly and a heat treatment. Each of these steps has a chance to induce supercrystalline defects, or potentially allow their migration and healing.

This work explores these aspects for inorganic-organic SCNCs via a combination of X-ray scattering and in-situ heating scanning transmission electron microscopy (TEM). Performing scattering experiments while rotating the sample by a small angular increment in a Small-Angle X-ray Scattering (SAXS) regime enables the determination of the full 3D reciprocal space map of the sample. If the sample is supercrystalline, it will display the Bragg peaks corresponding to its superlattice structure. Angular X-ray Cross-Correlation Analysis (AXCCA) of the 3D intensity distribution enables the identification not only of the average structure of the supercrystalline sample, but also of potential defects.[61–63] STEM, on the other hand, allows localized structural characterization at higher magnifications.[64]

We show here that, while all SCNCs feature predominantly face-centred cubic (*fcc*) superlattice arrangements, different types of lattice distortions and defects are detected in differently processed SCNCs. Pressing bulk SCNCs pellets leads to a "stretching" of the *fcc* superlattice into a slightly distorted triclinic one, whilst spherical supercrystalline supraparticles (SPs) show the presence of random hexagonal close-packed (*r-hcp*)[65,66] motifs within a prevalently *fcc* structure. Remarkably, heating the SCNCs at mild temperatures (up to 350 ºC) – such as those inducing organic crosslinking, and therefore even in the presence of a covalent network connecting the NPs – leads to the migration and healing of defects, as revealed for both superlattice stacking faults and inter-supercrystalline ("grain") boundaries.

**SCNCs nanostructure and mechanical properties**

The supercrystalline structure is obtained via self-assembly. To obtain macroscopic materials, self-assembly is conducted with two strategies: (1) solvent destabilization with the initial suspension of functionalized NPs in a die-punches assembly; (2) emulsion-templated self-assembly.[21,67] Method (1) yields bulk, mm-sized samples, which are subsequently pressed uniaxially to shape the material into pellets. This self-assembly method results in poly-supercrystalline materials, i.e. containing multiple superlattice domains with varying orientations, analogous to grains in polycrystalline materials. Method (2) yields a distribution of μm-sized supercrystalline spheres, i.e. supraparticles (SPs),[67,68] which can be used as building blocks for hierarchical bulk materials.[69] In all cases, the material building blocks are quasi-spherical (truncated cubooctahedral) iron oxide (magnetite, $Fe_3O_4$) NPs, with a radius of 7.4 ± 0.8 nm, surface-functionalized with oleic acid (Fraunhofer CAN GmbH, Hamburg, Germany). For the following analysis, samples with characteristic sizes in the μm range are considered, to assess no more than two supercrystalline domains at a time. From the bulk pellets, micropillars and lamellae are obtained via focused ion beam (FIB) milling, the former with a square cross-section to facilitate the subsequent X-ray analysis, and the latter for the scanning transmission electron microscopy (STEM) analysis. All details on SCNC processing are reported in previous publications,[5,21,67,70] and briefly summarized in the Methods and Supplementary Information (SI), section 1.

We therefore study two types of SCNCs: micropillars from bulk pellets (named "Pillars" in the following) and individual supraparticles ("SPs"). Fig. 1 shows SEM micrographs of the two types of SCNCs: Pillars with varying superlattice orientations in Fig. 1a-c and SPs in Fig. 1d-f. As the fracture surfaces of Fig. 1b,c show, in the bulk pellets surface ledges and terraces (step edges and kinks) can be found, along specific planes, due to the supercrystalline anisotropy. The presence of anisotropic ledges and terraces agrees with those previously detected at interfaces in conventional polycrystalline systems,[71–75] even though at a markedly larger length scale. In SPs, the superlattice orientation needs to be accommodated within the spherical shape. In line with previous reports, a finite number of particles (above 100) organizes into an icosahedral symmetry with either Mackay or anti-Mackay structures (Fig. 1e),[36,76] while for particle numbers above $10^6$, the structure transitions into single *fcc* domains (Fig. 1f).[35,36]

A key processing step in both cases is the heat treatment-induced crosslinking of the organic phase. In this system, crosslinking can be induced via heat treatment in a broad spectrum of temperatures and atmospheres.[25] In this study, we focus on heating for 18 min at 325 ºC in $N_2$ atmosphere, with a heating ramp of 1 ºC·min$^{-1}$, which is one of the most efficient and well-established routines.[21] Crosslinking of the organic phase leads to the transition from a material held together mainly by van der Waals interactions, to one in which the NPs are interconnected through a network of strong covalent bonds. This results in a significant increase in elastic modulus, hardness and strength.[5,21] It should nevertheless be mentioned that even before the heat-treatment, the SCNCs, both in bulk and in SP form, already show remarkably high mechanical properties (compressive and bending strength >100 MPa).[17]

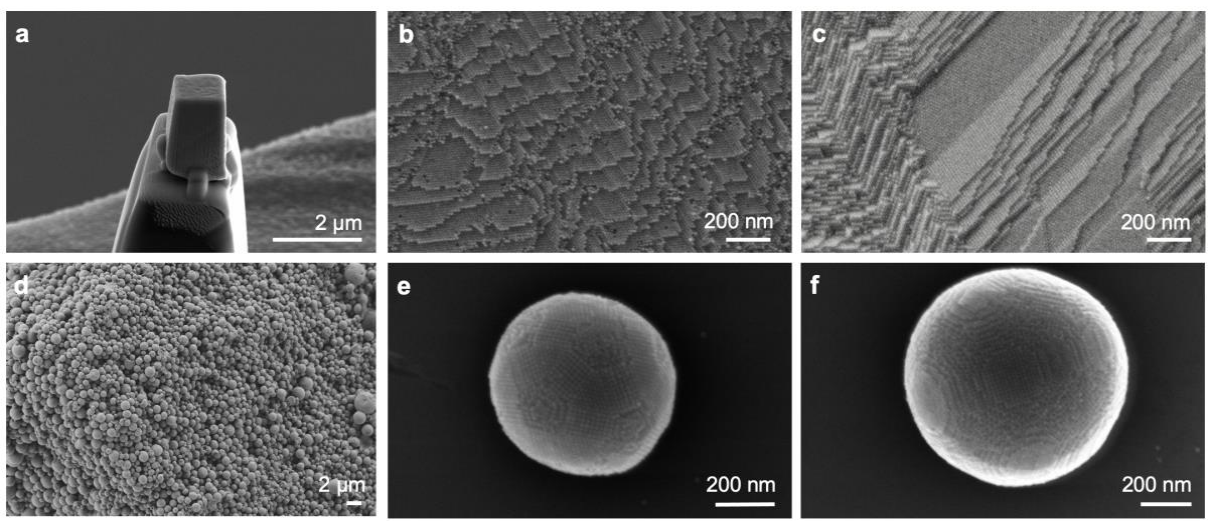

**Fig. 1: Nanostructure of the supercrystalline nanocomposites (SCNCs). a-c** Micropillar ("Pillar") from bulk material with representative nanostructures. **a** Pillar fixed on top of the pin used for the 3D X-ray analysis, **b,c** Secondary electron (SE) images of supercrystalline nanostructure from fracture surfaces of bulk SCNCs, with the organically-functionalized iron oxide NPs organized in periodic arrangements with multiple orientations, representative of the poly-supercrystalline character of the bulk samples. **d-f** SE images of supraparticles ("SPs") obtained via emulsion-templated self-assembly. **d** Overview of the gram-scale production of iron oxide-oleic acid SPs.[67] **e,f** Single SPs with evidence of (e) anti-Mackay superlattice in SP 2 and (f) single *fcc* arrangement of the functionalized NPs in SP 3.

Samples were measured by means of X-ray scattering, with a µm-sized focused synchrotron X-ray beam, before and after heat treatment. 3 Pillars and 3 SPs were analysed. Pillars 1 and 2 are tested once heat-treated ("HT"), while Pillar 3 is tested both before and after heat treatment. The Pillars have each a uniform superlattice orientation, i.e. consisting of a single supercrystalline domain. SPs 1 and 2 are heat-treated, while SP 3 is tested both before and after heat treatment. After the X-ray scattering experiment, the Pillars and SPs (in cross-linked state, HT) are also tested mechanically via microcompression. Finally, based on the findings emerged for the heat-treated materials via X-ray analysis, a grain boundary between supercrystalline domains, extracted in the bulk samples also used for the Pillars, is analysed via in-situ STEM.

The enhancement of the mechanical properties resulting from the organic crosslinking is confirmed by the microcompression tests. Data obtained from loading-unloading cycles during compression of Pillars and SPs (see Methods) show a mainly linear elastic deformation behaviour, with an average strength of ~500 MPa for the Pillars and an equivalent fracture strength of 300 MPa for the SPs (calculated as applied load divided by the equatorial cross-section of the sphere). All data is shown in SI section 2. Even though for SCNCs of analogous compositions even higher strength values have been reported,[5,21] these values are still remarkably high. Their slight decrease compared to previous studies [5,21] is attributed to the potential presence of damage or misalignments with respect to the applied load direction, due to the multiple micro-sample transfers and manipulations for the X-ray analysis and the mechanical tests.

**X-ray scattering analysis: Superlattice deformation with uniaxial pressing**

By means of rotation of the sample by 180° in the X-ray beam, we measured the scattered intensity in 3D reciprocal space for all samples (see SI section 3, Figs. S3-S8). An example of the scattering intensity distribution for sample Pillar 1 is shown in Fig. 2a, the corresponding distributions for other Pillars are shown in the SI, Figs. S4 and S6. It contains several Bragg peaks which by the relative angular positions can be attributed to an *fcc* structure in real space. The peaks with the lowest *q*-value belong to the $111_{fcc}$ peak family of an *fcc* structure. The corresponding directions of four of them are indicated in Fig. 2a. Using their angular positions, one can estimate the unit cell orientation in real space, as shown in Fig. 2b. In addition to the Bragg peaks, there is diffuse scattered intensity at lower *q*-values. The measured 3D diffraction patterns contain "flares" of intensity in the horizontal plane of reciprocal space that originate from scattering on the pillar walls. One should note also the splitting of the "flares", as the opposite pillar walls are not perfectly parallel. Their directions $n_1$, $n_2$ and $n_3$ coincide with the normal vectors to the pillar walls as shown in Fig. 2c. This allows to detect the orientation of the unit cell with respect to the pillar walls, as shown in Fig. 2b.

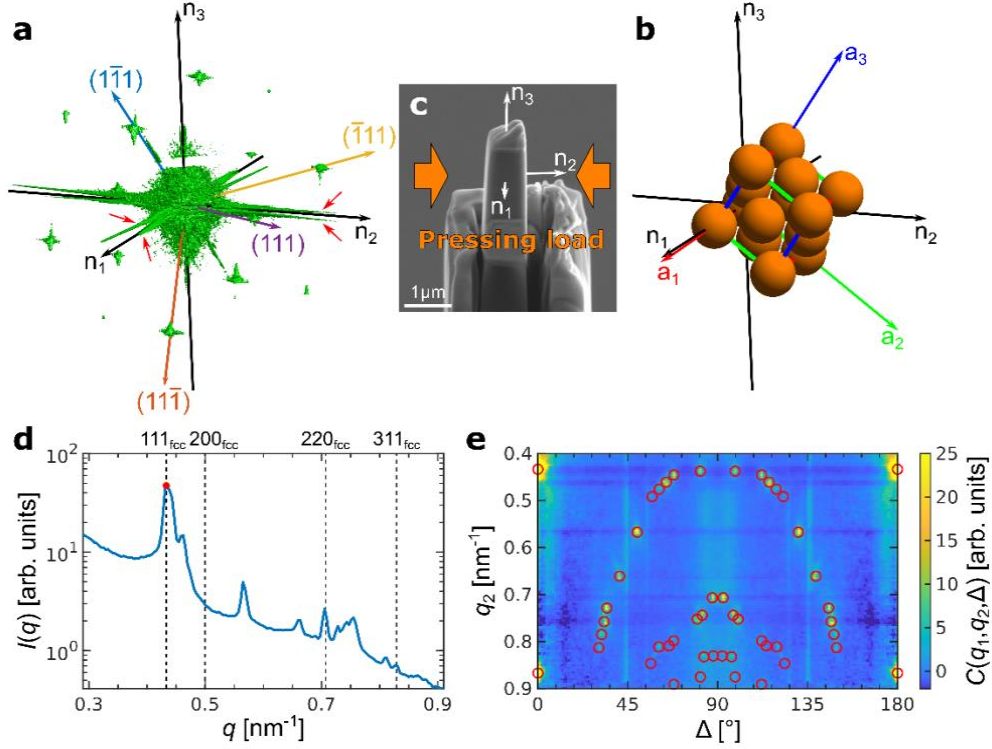

**Fig. 2: Angular X-ray Cross-Correlation Analysis (AXCCA) of SCNC Pillars. A** An isosurface of the measured scattered intensity distribution in 3D reciprocal space for Pillar 1. Four $(111)_{fcc}$ directions of the reciprocal lattice of a distorted *fcc* lattice are indicated by coloured arrows, as well as the normal vectors $n_1$, $n_2$ and $n_3$ to the pillar walls deduced from the intensity "flares" orientation. The red arrows indicated splitting of the intensity "flares" associated with the non-perfectly parallel pillar walls. **b** Orientation of a distorted *fcc* unit cell with respect to the pillar walls in real space. **c** A SEM image of the same pillar with indicated directions $n_1$, $n_2$ and $n_3$. The uniaxial stress direction applied during the sample preparation is also indicated. **d** Azimuthally averaged intensity profile of the 3D scattered intensity of Pillar 1, with the red point indicating $q_1 = 0.435$ nm$^{-1}$, used for the calculation of the cross-correlation functions (CCFs). The peak positions of an ideal *fcc* structure are indicated with vertical dashed lines. **e** CCFs $C(q_1,q_2,\Delta)$, calculated for $q_1$ (red point in **d**) and $q_2$ in the range of $0.4 - 0.9$ nm$^{-1}$, stacked along the vertical axis $q_2$, with the peak positions for the optimized unit cell parameters marked by red circles.

However, the average azimuthal profile of the scattered intensity distribution, shown in Fig. 2d, does not correspond perfectly to an *fcc* superlattice and contains many additional peaks. To extract the unit cell parameters, we applied the AXCCA technique to the collected 3D X-ray scattering data (see Methods and Ref. 60 for details).[62] It is assumed that the superlattice has the lower-symmetry primitive triclinic structure with parameters *a'*, *b'*, *c'* and angles between them *α'*, *β'*, *γ'*. The peak at $q_1 \approx 0.435$ nm$^{-1}$ is attributed to the 100 reflection (primitive unit cell). The Cross-Correlation Functions (CCFs) are then calculated for the intensity taken at this $q_1$ and all other $q_2$ momentum transfer values in the range of 0.4 – 0.9 nm$^{-1}$. The resulting CCFs $C(q_1,q_2,\Delta)$ are shown in Fig. 2e. There are several peaks at different $q_2$ values and different angles $\Delta$, highlighted with red circles in the same Fig. 2e. Each peak corresponds to a vector $\mathbf{g_2}$ of the reciprocal lattice that has length $q_2 = |\mathbf{g_2}|$ and relative angle $\Delta$ with respect to the vector $\mathbf{g_1} = \mathbf{b_1}$ with the length $q_1 = |\mathbf{g_1}| \approx 0.435$ nm$^{-1}$. The unit cell parameters are obtained by optimization to fit the experimental peak positions, with the mean value of the experimental CCFs at the calculated peak positions used as a metric as shown in SI section 3.

The superlattice unit cell parameters of all tested pillars are given in Table 1. It is immediately noticeable that all values are very close to those of an *fcc* structure (for which $a' = b' = c'$ and $\alpha' = \beta' = \gamma' = 60°$), but yet well separated from them, supporting the description of the lattice as a "distorted *fcc*" with the parameters *a*, *b*, *c* and *α, β, γ*, which are also given in Table 1.

**Table 1.** Unit cell parameters of the Pillar samples extracted by Angular X-ray Cross-Correlation Analysis (AXCCA).

| Sample | Primitive unit cell parameters | | | | | | Face-centered unit cell parameters | | | | | |
|---|---|---|---|---|---|---|---|---|---|---|---|---|
| | $a'$, nm | $b'$, nm | $c'$, nm | $\alpha'$, ° | $\beta'$, ° | $\gamma'$, ° | $a$, nm | $b$, nm | $c$, nm | $\alpha$, ° | $\beta$, ° | $\gamma$, ° |
| Pillar 1 (HT) | 16.8 ±0.2 | 17.0 ±0.5 | 17.8 ±0.3 | 58.7 ±1.8 | 60.8 ±0.7 | 68.5 ±1.1 | 22.2 ±0.3 | 26.5 ±0.7 | 25.6 ±0.6 | 93.8 ±2.4 | 91.1 ±2.0 | 91.7 ±1.8 |
| Pillar 2 (HT) | 17.1 ±0.2 | 17.4 ±0.3 | 17.8 ±0.4 | 57.0 ±1.6 | 63.5 ±0.8 | 60.9 ±0.8 | 23.6 ±0.4 | 26.2 ±0.4 | 24.3 ±0.4 | 93.3 ±1.5 | 88.9 ±1.6 | 89.0 ±1.5 |
| Pillar 3 (before HT) | 18.2 ±0.5 | 16.9 ±0.5 | 17.0 ±0.3 | 61.3 ±1.8 | 58.1 ±0.6 | 61.3 ±1.5 | 24.5 ±0.8 | 23.7 ±0.9 | 25.8 ±0.6 | 93.1 ±2.5 | 86.8 ±2.3 | 91.1 ±2.0 |
| Pillar 3 (after HT) | 18.0 ±0.5 | 16.8 ±0.4 | 16.9 ±0.4 | 61.5 ±1.9 | 58.4 ±0.8 | 61.6 ±1.5 | 24.4 ±1.0 | 23.7 ±0.8 | 25.6 ±0.7 | 93.1 ±2.4 | 87.0 ±2.7 | 91.1 ±2.6 |

Interestingly, a pattern emerges: all micropillars have one primitive unit cell parameter close to 18 nm, while the other two are closer to 17 nm, and an angle smaller than 60º, with the other two being larger instead. This suggests a consistency in the superlattice distortion mechanism. Previous studies have shown that the uniaxial pressing step (with confinement in a rigid die) can lead to superlattice anisotropies.[18] Here, the analysis of the orientation of the superlattice within each micropillar and with respect to the applied pressing load confirms this effect, see Fig. 2b-c. This is especially clear in two Pillars (1 and 2), which show the same superlattice orientation, and more specifically the $[0\bar{1}1]_{fcc}$ axis oriented parallel to the vertical axis of the pillars, and the $[100]_{fcc}$ and $[011]_{fcc}$ axes in the pillars' cross-sectional plane. The superlattice has the shortest unit cell constant, and is thus compressed most, along the $[100]_{fcc}$ axis, and secondarily along the $[011]_{fcc}$ one. Both these axes lay in the cross-sectional plane of the pillar, while the dimension along the pillars' vertical axis is the largest one. One should also note that the largest angle $α$ is opening towards the $[011]_{fcc}$ direction. Since the pillars were extracted from an axial cross-section of the bulk cylindrical pellets, the vertical direction of each pillar is oriented perpendicular to the pressing load, thus confirming that the pressing step leads to a superlattice distortion. The superlattice is stretched in the direction perpendicular to the pressing load. Indeed, the SCNCs have been shown to allow for compaction and superlattice stretching due to the presence of free volume, detected in previous studies via both TEM and positron annihilation lifetime spectroscopy (PALS).[54,77]

Pillar 3 has, instead, a superlattice orientation that leads to a less evident effect of the pressing step, although even there the angles $α$ and $γ$ opening in the horizontal plane are bigger than $β$ opening towards the vertical direction (see SI, Fig. S6). This pillar was, additionally, analysed via X-rays both before and after the crosslinking-inducing heat treatment (HT). It emerges that, even though the organic phase undergoes significant changes and is partially removed during this step,[25,54] the superlattice does not shrink significantly. As shown in Table 1, the superlattice parameters show only a very slight decrease in their average value, while the average angle values show a small increase, leading to negligible changes given the measurement error.

**X-ray scattering analysis: Stacking faults in supraparticles**

The SPs feature an undistorted *fcc* superlattice, since they do not undergo a pressing step. However, the measured X-ray scattering intensity distribution, shown in Fig. 3a for the sample SP 1, contains not only the corresponding Bragg peaks as expected. There are also rod-like features known as Bragg rods,[78] which indicate here the presence of stacking faults in the structure. Note that while these kinds of rods could also potentially emerge from the tested samples' planar truncation, such as the Pillars' walls, they are also detected in the SPs, which have minor levels of truncated areas, and hence we can conclude that stacking faults are the main cause leading to their appearance. For close-packed crystals, the Bragg rods are continuous intensity modulations along the $h \cdot \mathbf{b_1} + k \cdot \mathbf{b_2} + l \cdot \mathbf{b_3}$ lines, where $h - k \neq 3n; h, k, n \in Z$ and $l \in R$, and $\mathbf{b_1}$, $\mathbf{b_2}$, $\mathbf{b_3}$ are the reciprocal basis vectors of the corresponding *hcp* lattice. The intensity modulation along them is dependent on the stacking order.

The corresponding average radial profile shown in Fig. 3b contains two bright peaks, which can be attributed to the $111_{fcc}$ and $220_{fcc}$ reflections of the *fcc* structure. An additional intensity on the left side of the $111_{fcc}$ reflection can be attributed to the $100_{hcp}$ reflection of an *hcp* structure with the same nearest neighbour distance as expected for a close-packed structure with stacking faults. Small peaks between the $111_{fcc}$ and $220_{fcc}$ reflections originate from the intensity distribution along the Bragg rods. Thus, the structure of the SPs can be characterized as random hexagonal close-packed (*r-hcp*)[65,66] structure containing both *fcc* and *hcp* stacking motifs. For SP 3, we observe a similar 3D intensity distribution and azimuthal profile, as shown in Fig. 4a,b.

To refine the identification of the SP structure, we applied AXCCA to the 3D intensity distribution. The CCFs were calculated for the intensities taken at $q_1$ corresponding to the $111_{fcc}$ reflection and all other momentum transfer values $q_2$ in the range of $0.4 - 0.9$ nm$^{-1}$. The resulting CCFs are shown in Fig. 3c and 4d. Differently from the Pillars, the maps for the SPs contain not only peaks, but also "arcs" of intensity (in yellow and marked with red dashed lines in the figures). They originate from the correlation between the $111_{fcc}$ reflections and the Bragg rods of the $10l_{hcp}$ family.

The SP unit cell parameters were optimized with the same procedure as for the triclinic structure of the Pillars, even though in this case, in absence of superlattice distortion, there is only one parameter to be determined: the nearest neighbour distance $d_{NN}$ between adjacent NPs. The optimized $d_{NN}$ values are summarized in Table 2. One can calculate the unit cell parameters $a_{fcc} = \sqrt{2} d_{NN}$ for the fcc and $a_{hcp} = d_{NN}, c_{hcp} = \sqrt{8/3}\, d_{NN}$ for the hcp structures from the $d_{NN}$ values.[28] The nearest neighbour distance is consistently ~16 nm, again showing no detectable superlattice shrinkage associated with the heat treatment. For the hcp structure, both $100_{hcp}$ and $002_{hcp}$ peaks are detected and no deviations from the ideal ratio are observed. Based on these features, the structure of the SPs can be described as an r-hcp with prevalent fcc stacking motifs. No peaks that can be attributed solely to hcp domains are observed, but due to the stacking faults there can be hcp areas with the thickness of a few NP layers.

**Table 2.** Nearest-neighbour distances and unit cell parameters of SPs as extracted by AXCCA.

| Sample | Nearest-neighbor distance $d_{NN}$, nm | fcc unit cell parameter $a_{fcc}$, nm | hcp unit cell parameter $a_{hcp}$, nm | hcp unit cell parameter $c_{hcp}$, nm |
|---|---|---|---|---|
| SP 1 (HT) | 16.0±0.1 | 22.6±0.2 | 16.0±0.1 | 26.1±0.2 |
| SP 2 (HT) | 16.2±0.2 | 22.9±0.3 | - | - |
| SP 3 (before HT) | 15.9±0.2 | 22.5±0.3 | 15.9±0.2 | 26.0±0.3 |
| SP 3 (after HT) | 15.9±0.2 | 22.5±0.3 | 15.9±0.2 | 26.0±0.3 |

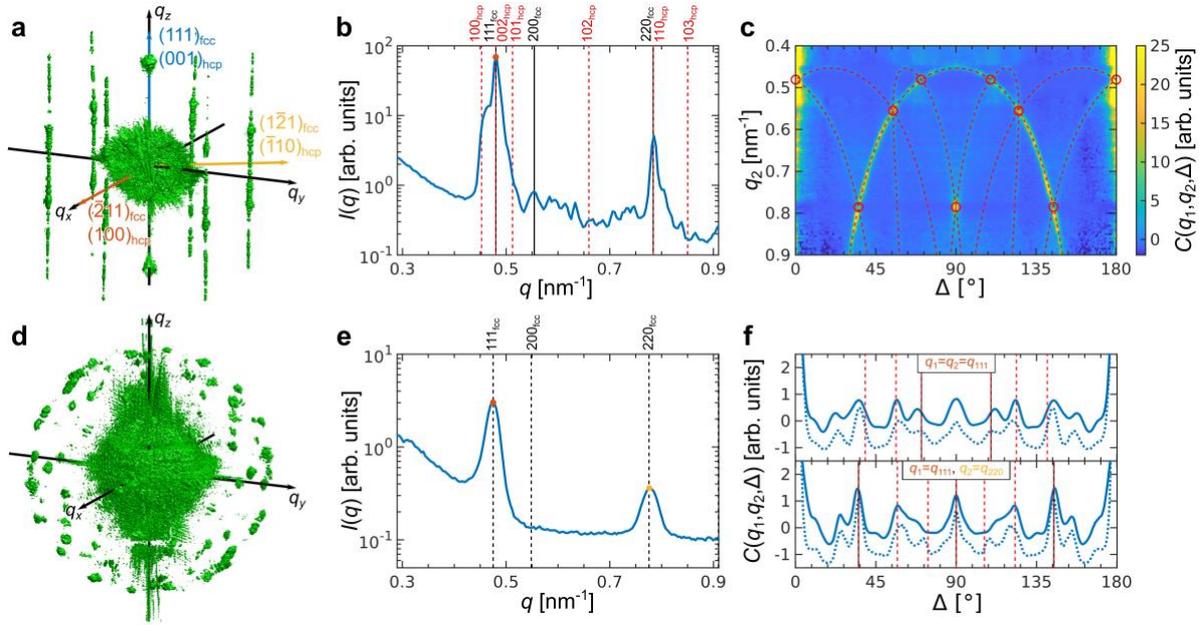

**Fig. 3: Angular X-ray Cross-Correlation Analysis (AXCCA) of supraparticles (SPs). a** An isosurface of the measured scattered intensity distribution in 3D reciprocal space for SP 1. The $(111)_{fcc}/(001)_{hcp}$ reciprocal direction corresponding to the stacking direction as well as the $(\bar{2}11)_{fcc}/(100)_{hcp}$ and $(1\bar{2}1)_{fcc}/(\bar{1}10)_{hcp}$ reciprocal directions constituting the basis in hexagonal close-packed planes are shown with coloured arrows. The Bragg rods are an indication of the presence of stacking faults. **b** Averaged radial profile of the 3D scattered intensity of SP 1, with the red point indicating $q_1 = 0.480$ nm$^{-1}$, used for the calculation of the CCFs. The peak positions corresponding to $fcc$ and $hcp$ structures are indicated with black and red dashed lines, respectively. The SP structure is thus a random $hcp$ ($r$-$hcp$)[65,66] containing both $fcc$ and $hcp$ stacking motifs. **c** CCFs $C(q_1,q_2,\Delta)$ calculated for $q_1$ indicated in (**a**) and $q_2$ in the range of 0.4 – 0.9 nm$^{-1}$. The CCFs are shown as a heat map stacked along the vertical axis $q_2$. The peak positions for the optimized nearest-neighbour distance $d_{NN}$ corresponding to the maximum of correlation are marked with red circles for the $fcc$ structure, and the red dashed lines for the correlations with the Bragg rods from an $r$-$hcp$ structure. The "arcs" of intensity originate from the correlation between the $111_{fcc}$ reflections and the Bragg rods of the $10l_{hcp}$ family. **d** An isosurface of the measured scattered intensity distribution in the 3D reciprocal space for SP 2. **e** Averaged radial profile of the 3D scattered intensity of SP 2, with the red point indicating $q_{111} = 0.475$ nm$^{-1}$ and the yellow point indicating $q_{220} = 0.775$ nm$^{-1}$, used for the calculation of the CCFs. The peak positions for an $fcc$ structure are indicated with blacked dashed lines. **f** CCFs $C(q_1,q_2,\Delta)$ calculated for $q_1 = q_2 = q_{111} = 0.475$ nm$^{-1}$ (top) and $q_1 = q_{111} = 0.475$ nm$^{-1}$, $q_2 = q_{220} = 0.775$ nm$^{-1}$ (bottom). The solid lines are calculated for the experimental intensity distribution and the dashed ones for the simulated. Several peaks here cannot be attributed to a single $fcc$ lattice, and indeed SP 2 is found to have an anti-Mackay structure (see Fig. 1e, SI Section 4 and Fig. S11).

SP 2 has a very distinctive structure. The 3D intensity distribution shown in Fig. 3d contains many Bragg peaks at the same $q$-value, which cannot be attributed to a single crystalline structure. On the other hand, the radial profile shown in Fig. 3e contains two peaks at $q$ = 0.475 nm$^{-1}$ and 0.775 nm$^{-1}$ that can be attributed to 111$_{fcc}$ and 220$_{fcc}$ reflections of an *fcc* structure with $a_{fcc}$ = 22.9 ± 0.5 nm ($d_{NN}$ = 16.2 ± 0.3 nm). The CCFs calculated for these two *q* values are shown in Fig. 3f. They contain several peaks that cannot be attributed to a single *fcc* lattice. Given its size, this supraparticle is indeed expected to have the so-called anti-Mackay structure,[76] which consists of many mutually twinned *fcc* domains, also in line with SEM observations (see Fig. 1e). The geometric calculation of the expected correlation peak positions for the anti-Mackay structure is quite cumbersome, due to the complex relative orientations of the multiple twinned domains. Instead, we simulated the 3D scattered intensity distribution for a SP with the anti-Mackay structure and similar size as described in SI, Section 4. Comparison of the CCFs calculated for the simulated intensity distribution with the experimental ones shown in Fig. 3f confirms the anti-Mackay structure of SP 2. We refer to Supplementary Information, Fig. S11 for the full 2D experimental and simulated CCF maps.

### Thermal annealing of planar defects

The heat treatment has been shown to induce no detectable shrinkage for both Pillars and SPs (see Table 1 and Table 2). However, it plays an important role in the mobility, rearrangement and migration of superlattice defects. We show in the following how annealing the SPs leads to the healing of the stacking faults, while in the bulk samples heating leads to rearrangement of a supercrystalline grain boundary.

The scattered intensity distribution of SP 3 before heating, shown in Fig. 4a, is similar to that of SP 1 (Fig. 3a). It contains Bragg peaks and Bragg rods and can also be characterized as an *r-hcp* structure with *fcc* and *hcp* stacking motifs and many stacking faults. The radial profile shown in Fig. 4b contains the corresponding Bragg peaks.

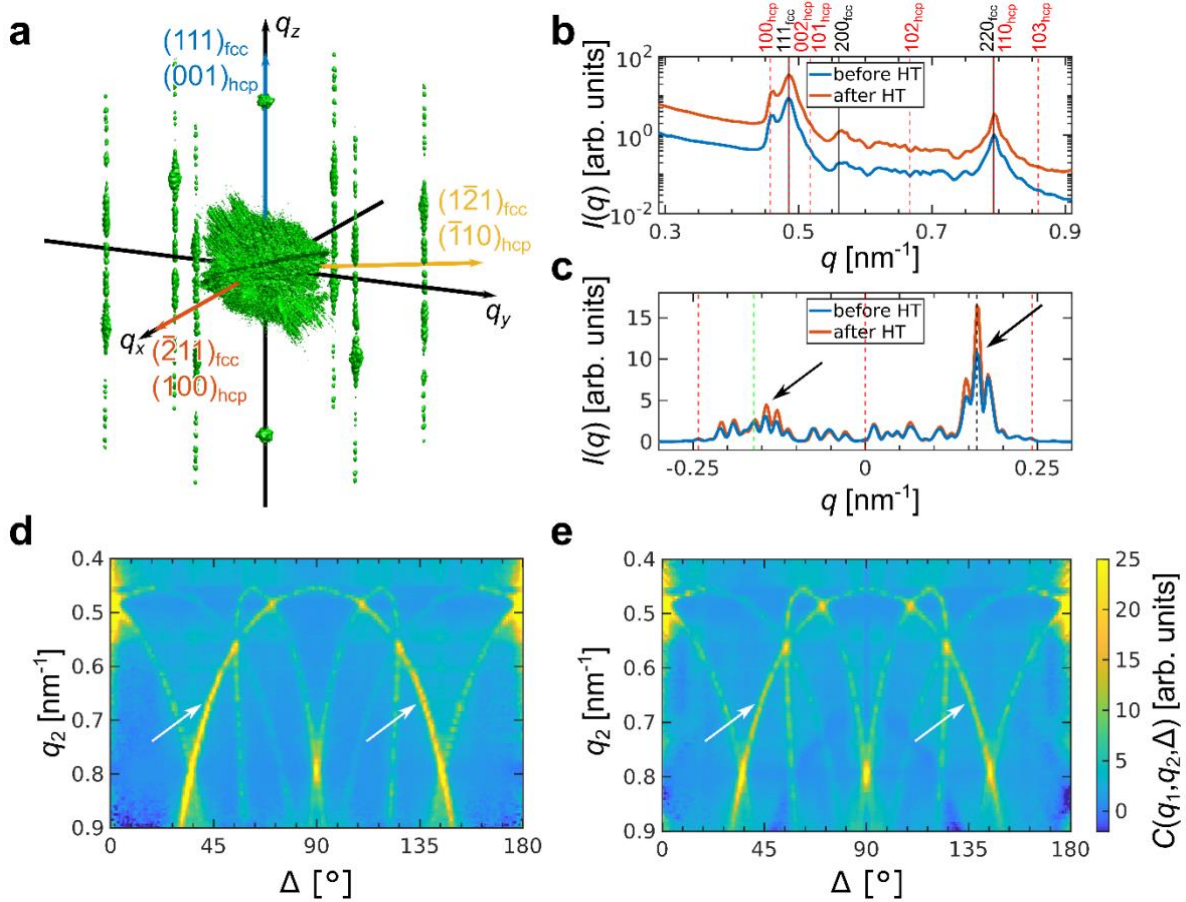

**Fig. 4: Angular X-ray Cross-Correlation Analysis (AXCCA) of a SP before and after heat treatment. a** An isosurface of the measured scattered intensity distribution in 3D reciprocal space for SP 3. The $(111)_{fcc}/(001)_{hcp}$ reciprocal direction corresponding to the stacking direction as well as the $(\bar{2}11)_{fcc}/(100)_{hcp}$ and $(1\bar{2}1)_{fcc}/(\bar{1}10)_{hcp}$ reciprocal directions constituting the basis in hexagonal close-packed planes are shown with coloured arrows. The Bragg rods are an indication of the presence of stacking faults. **b** Averaged radial profiles of the 3D scattered intensities of SP 3 before (blue line) and after (red line) heat treatment. The profiles are shifted vertically for clarity. The peak positions corresponding to *fcc* and *hcp* structures are indicated with black and red dashed lines, respectively. **c** Averaged intensity profile along the $10l_{hcp}$ Bragg rods before (blue line) and after (red line) heat treatment. The peak positions corresponding to *fcc* and *hcp* structures are indicated with black and red dashed lines, respectively, and the twinned *fcc* structure is indicated with green dashed lines. **d, e** CCFs $C(q_1,q_2,\Delta)$ calculated for $q_1$ corresponding to the $111_{fcc}$ Bragg peak and $q_2$ in the range of 0.4 – 0.9 nm$^{-1}$ before (**d**) and after (**e**) heat treatment. The CCFs are shown as a heat map stacked along vertical axis $q_2$. **c**, **d** and **e** point at the thermal annealing of the stacking faults, via the intensity profiles along the $10l_{hcp}$ Bragg rods shown in **c** and the lower intensity of the central yellow "arc" in **e** (after heat treatment) compared to **d**, as marked by the arrows. Note that the dark blue arc in **e** is an artefact originating from the fact that we do not measure the whole reciprocal space.

Remarkably, the heat treatment does not lead to major changes in the 3D intensity distribution and the azimuthal profile. However, the heat treatment does lead to the redistribution of intensities in the CCF maps of the same SP 3 before and after heat treatment (Fig. 4d,e). One can see that the central "arc" has higher intensity before heat treatment than afterwards. This "arc" comes from the correlations of the $10l_{hcp}$ Bragg rods with the $111_{fcc}/002_{hcp}$ stacking-independent Bragg peaks that are normal to the hexagonal planes. Other "arcs" originate from the correlations between the Bragg rods and the $111_{fcc}$ peaks that are not normal to the hexagonal planes and thus originate purely from *fcc* motifs. The intensity of the stacking-independent peaks is supposed to be constant, but the CCFs are normalized by the total intensity. It thus emerges that the intensities of the stacking-dependent $111_{fcc}$ peaks have raised upon heat treatment, indicating an increased ratio of the *fcc* / *hcp* domains. This effect is also confirmed by higher intensities at the peak positions that are characteristic for an *fcc* structure in the CCF maps, and it can be seen in the intensity profiles along the $10l_{hcp}$ Bragg rods shown in Fig. 4c. Thus, we conclude that the stacking faults are healed by means of heat treatment.

To verify this effect on the stacking faults, the temperature-dependent evolution of a SCNC with *hcp* and *fcc* domains was studied using all-atom molecular dynamics (MD) simulations. The simulated SCNC consists of magnetite NPs, which are 4 nm in diameter and functionalized with oleic acid molecules. These have been used to build and equilibrate an *fcc* SCNC. From the equilibrated *fcc* SCNC, a "ABCABCABABAB" system was generated. This represents the shortest sequence that hosts both *fcc* and *hcp*, while allowing in principle for a transition to a pure *fcc* or *hcp* crystal structure. The generated SCNC contains 24 functionalized NPs with approximately 400,000 atoms, see Fig. 5a.

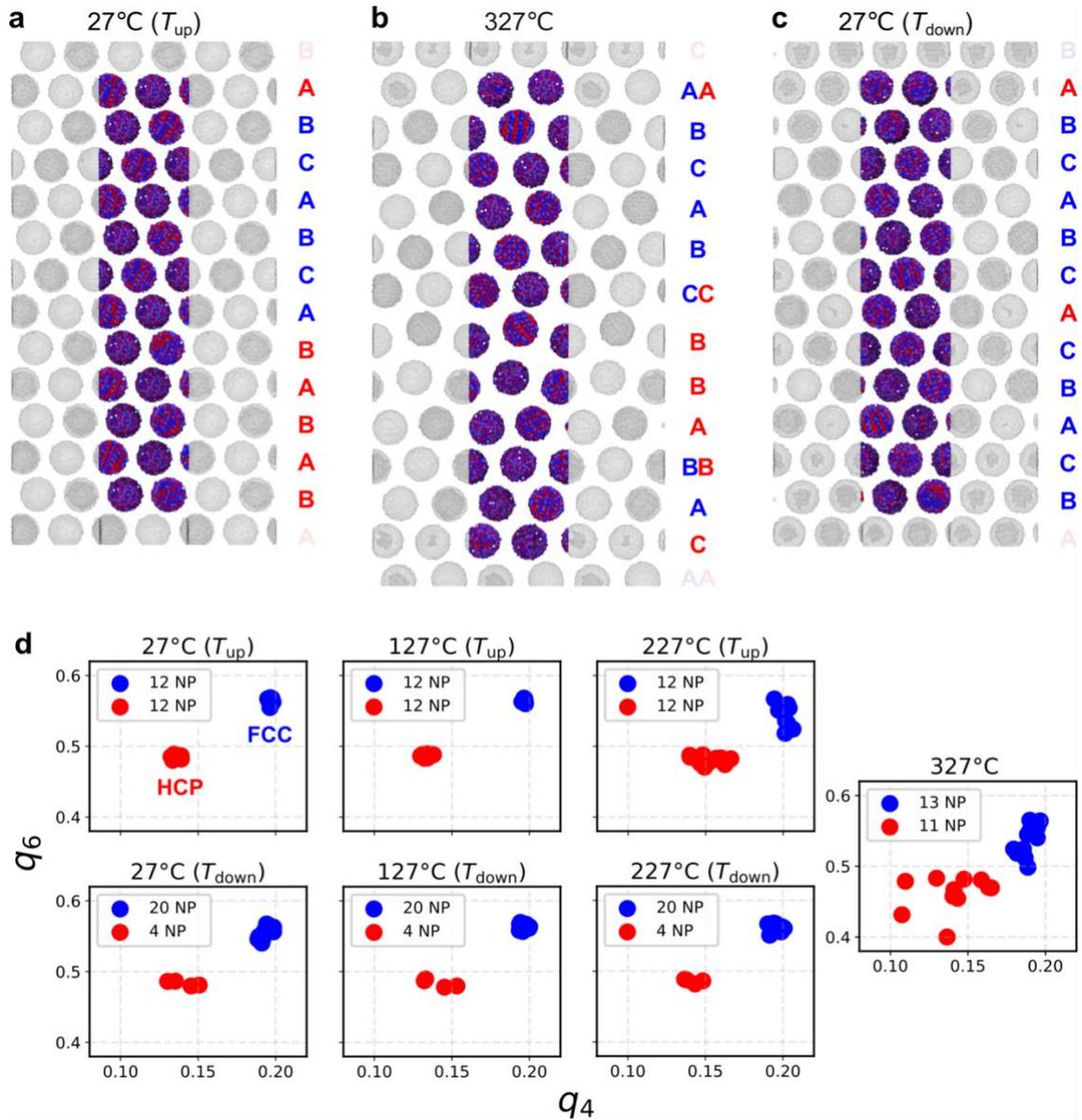

**Fig. 5: All-atom simulation of the temperature-dependent evolution of a *hcp/fcc* supercrystalline nanocomposite (SCNC).** The temperature and whether the system is heated ($T_{up}$) or cooled ($T_{down}$) are specified in each plot title. **a** Structure at 27 °C ($T_{up}$); **b** at 327 °C; and **c** at 27 °C ($T_{down}$). The stacking sequence with the corresponding structure's colour code is shown on the right. For clarity of representation, the oleic acid molecules are not shown. The simulation boxes are highlighted, and each layer contains two NPs. **d** Steinhardt bond order parameters, $q_4$ and $q_6$, of the system containing 24 oleic acid functionalized magnetite NPs. *fcc*-like and *hcp*-like NPs are indicated by blue and red dots, respectively. The amount of the corresponding NPs is given in the legend.

The system was heated for 75 ns from 27 °C to 327 °C, denoted by $T_{up}$, at a constant pressure of 1 atm in an *NpT* ensemble. Next, the system was held at 327 °C for 10 ns and then cooled down to 27 °C, denoted by $T_{down}$, over 75 ns. Steinhardt bond order parameters[79], particularly, the $q_4$ and $q_6$ parameters, were employed to determine the local supercrystalline structure of the NPs, namely for distinguishing between *fcc* and *hcp* local ordering. At 27 °C ($T_{up}$), $q_4$ and $q_6$ bond order parameters indicate that 12 NPs are in a *fcc* and 12 in a *hcp* configuration (see Fig. 5d). Both *fcc* and *hcp* NPs remain relatively stable close to 327 °C. At around 327 °C, the distances between NPs have increased, the *hcp* NPs start to reorganize, while the *fcc* NPs remain in their initial configuration. After cooling down to 27 °C ($T_{down}$), the reorganization of *hcp* NPs becomes more apparent, with 8 out of 12 transitioning to an *fcc* structure. A closer look at the system at 27 °C ($T_{down}$) (see Fig. 5c) reveals that the *hcp* signal in the $q_4$-$q_6$ analysis stems from twin boundaries in the *fcc* crystal structure, resulting in the stacking order of "(A)BCABC(A)CBACB", where twin boundaries are highlighted with brackets.

These results show that at the applied temperature the SCNC can reorganise from an *hcp* arrangement into *fcc*, in very good agreement with the experimental observations. It should be noted that due to the relatively small size of the simulated system, containing only 24 NPs, finite size effects in the simulations cannot be ruled out completely which might affect the presence and particularly the density of defects such as twin boundaries.

In general, the occurrence of stacking faults and twin boundaries in SCNC can be expected and was observed in iron oxide–oleic acid bulk SCNC.[5] Remarkably, however, their rearrangement and healing – to the best of the authors' knowledge – had not yet been detected. Annealing of twins and their migration is a well-known phenomenon, even though not fully understood, in polycrystalline metallic systems at temperatures above the recrystallization ones,[80] and the healing of stacking faults has also been observed in several crystalline materials.[81,82] Here, however, we observe healing of these planar defects, present in the supercrystals as an outcome of the SP formation, at significantly larger length-scales. The transition from *hcp* to *fcc* motifs has been predicted in colloidal crystals of hard spheres, in months-to-years timeframes.[83] What stands out here is the combined experimental and numerical observation of this transition happening in supercrystals containing organic ligands, within hours, and during a heat treatment that also leads to the crosslinking of the organic ligands, which induces a multi-fold strengthening of the SCNCs.

In the bulk samples from which the Pillars have been extracted, 2D defects are also expected, in the form of stacking faults and inter-supercrystalline, "grain", boundaries.[54] The temperature-induced migration of these kinds of 2D defects is then demonstrated with an in-situ STEM heating experiment on a supercrystalline grain boundary from a bulk SCNC. A thin lamella was obtained via FIB milling from an area that included such a boundary in a non-HT bulk SCNC. The boundary was oriented close to edge-on condition, such that the defects along it could be discerned. Fig. 6 presents high angle annular dark field (HAADF) STEM micrographs, demonstrating disconnections along the boundary that is oriented close to edge-on condition (see also SI section 6 and Fig. S14). Disconnections are line defects that occur along interfaces and that have a step and/or a dislocation component,[84–89] and they have so far been detected in polycrystalline materials, such as at general grain boundaries in $SrTiO_3$.[71–73] and at grain and phase boundaries in $\beta$-$Yb_2Si_2O_7$.[74,75] Anisotropic motion of disconnections has also been detected during in-situ STEM heating experiments, confirming that such motion is the mechanism of grain boundary migration in both general and high-symmetry grain boundaries.[87]

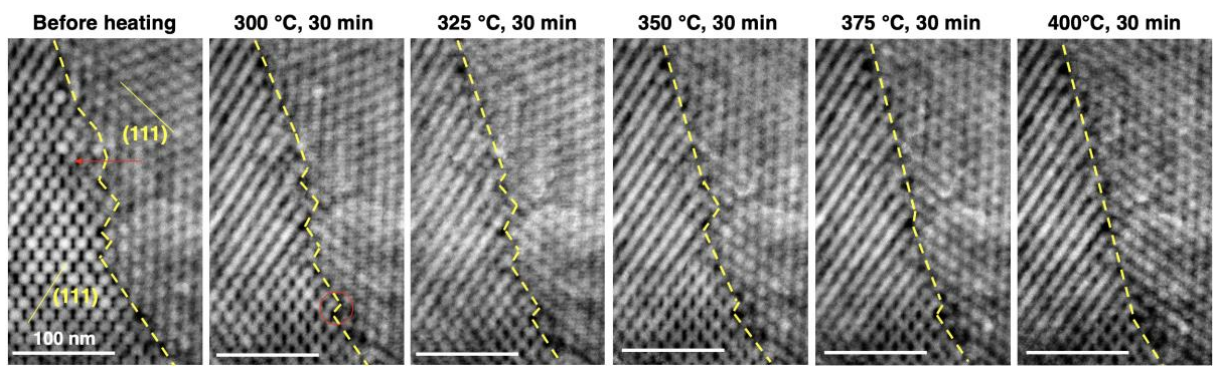

**Fig. 6: HAADF STEM micrographs of an in-situ heating experiment of a supercrstalline grain boundary.** Disconnections with grain boundary and step planes parallel to $\{111\}_{fcc}$ planes are demonstrated in bulk SCNCs, and disconnection migration is noted upon heating. The area is observed edge-on with zone axis $[0\bar{1}1]_{fcc}$ (see also Fig. S14). The dashed yellow lines mark the grain boundary and step planes.

Here, the boundary and step planes are parallel to specific crystallographic planes, i.e. $\{111\}_{fcc}$ superlattice planes, throughout the heating experiment, indicating their anisotropy in the case of this boundary. Even though we consider here a single general supercrystalline grain boundary, the anisotropy of the ledges and terraces at the fracture surfaces of SCNCs (Fig. 1), as well as the anisotropy of disconnections and ledges and terraces at interfaces in various conventional polycrystalline systems, suggest that other supercrystalline grain boundaries would exhibit the same anisotropy.

Most importantly, upon heating the disconnections migrate along the boundary, resulting in a decrease in the local number of steps. Whether the change in the number of steps is local or reduced throughout the entire boundary, the step height appears to remain similar, such that step bunching is not noted. These results extend the concept of anisotropic disconnection motion being the mechanism of grain boundary migration, and they provide experimental evidence well beyond previously studied polycrystalline systems,[71,72,85–87] i.e. in SCNCs with a lattice length-scale two orders of magnitude larger than those of conventional crystals, and on longer timescales (minutes instead of seconds). Given the anisotropy of the ledges and terraces in SCNCs, as well as anisotropy of disconnections and ledges and terraces at interfaces in polycrystalline systems, it is expected that other general supercrystalline grain boundaries would indicate similar anisotropic behaviour of disconnections.

Note that a very significant degradation of the organic phase is typically observed at temperatures above 350 ºC,[17,21] and thus NP sintering is expected in the last in-situ heating stages (at 375 and 400 ºC). On the other hand, the TEM lamella is fixed at its edges for the in-situ study, hampering the shrinkage. Fig. S14 shows the estimated distance between close-packed $\{111\}_{fcc}$ superlattice planes with increasing temperature, both from the local TEM measurement and from global SAXS data on the whole bulk SCNC from which the lamella is extracted. For the TEM case, since both abutting grains are not fully aligned in zone axis, but rather near it, and minor sample tilts as well as sample bending occur during the experiment, the interplanar distances cannot be determined with high precision, but only estimated. The threshold value below which NPs start to sinter (12.2 nm) is obtained with the assumption that the oleic acid on the NP surfaces is completely removed and the NPs thus become in contact with each other. In all cases, one can see that the interplanar distance remains above the threshold value corresponding to the onset of necking and sintering. This might suggest that, additional to the lamella constraint, the high symmetry of the NP *fcc* arrangement and the quasi-constant NP size prevent sintering, even after the pinning effect of the organic ligands is removed.[90]

**Conclusions**

This study provides insights into how distinct processing stages - self-assembly, pressing, and heat treatment - affect the formation, evolution, and healing of supercrystalline defects in high-strength inorganic-organic supercrystalline nanocomposites (SCNCs). The combination of AXCCA and in-situ STEM reveal how superlattice distortion, stacking faults, and supercrystalline grain boundaries all contribute to the structural complexity of SCNCs. In bulk pellets, the pressing step leads to a distortion of the otherwise *fcc* superlattice, which becomes stretched in the plane perpendicular to the applied load. Supraparticles obtained via emulsion-templated self-assembly, instead, show *fcc* NP arrangements with the presence of stacking faults and anti-Mackay structures, depending on their size.

Remarkably, we show that annealing at crosslinking-relevant temperatures (~350 °C) not only strengthens the material but also drives the migration and healing of planar defects. The stacking faults are found to be partially removed, as also confirmed via MD simulations, while an inter-supercrystalline grain boundary is found to become mobile, along which anisotropic disconnection motion is noted. The heat treatment, typically applied to strengthen the SCNCs, is then found to serve the additional role of inducing defect migration and healing.

The demonstrated thermally-activated migration of planar defects stands out in the analysis of supercrystals, because it does not occur during self-assembly or in low-viscosity films, but in a material state that cannot be considered as soft matter. Even before heat treatment these SCNCs reach bending and compressive strengths beyond 100 MPa. The parallelism with defect migration in crystalline materials becomes thus significantly more apparent and supported by both numerical and experimental evidence. We envision a plethora of new insights on defect migration in crystals based on supercrystalline platforms, and additionally the establishment of new design concepts for defect engineering in NP assemblies according to a "thermally programmable supercrystallinity" concept.

**Methods**

**Samples processing**

The SCNCs building blocks are iron oxide NPs that are surface-functionalized with oleic acid (Fraunhofer CAN GmbH, Hamburg, Germany). The size of the NPs has been assessed via small angle X-ray scattering (SAXS) of the initial NP suspension in toluene, according to a method reported elsewhere.[25,91] The NP radius is determined as 7.4 ± 0.8 nm. This is considered to be the radius of the inorganic core, without the organic functionalization, since the latter is markedly less detectable via X-rays, especially when the NPs are still in suspension.

The bulk SCNCs are prepared via self-assembly by the solvent destabilization method.[21] The NP suspension, with a concentration of 40 g·L$^{-1}$, is filled into an assembly of die and punch with a 14 mm diameter cavity, so that the pressing step can immediately follow the self-assembly. It is then placed in a desiccator, where the atmosphere is enriched with ethanol, which serves as destabilization agent when it slowly diffuses into the suspension. The process lasts approximately 15 days. The self-assembled samples, which at this point are sedimented at the bottom of the die-punch assembly, are recovered by removing the supernatant with a pipette. The SCNCs so obtained are dried for 24 h at ambient conditions and 2 h in vacuum. The pressing step finally follows, by applying 50 MPa via a second punch (in the rigid die) at a temperature of 150 ºC. The bulk cylindrical SCNC pellets are 14 mm in diameter and ~4 mm thick.

The supraparticles were prepared via emulsion-templated self-assembly. A solution of 113 g polyoxyethylene (20) sorbitan monolaurate (Tween™ 20, Appli Chem, ≥100%), 3.48 g sorbitan monolaurate (Span™ 20, Merck, synthesis grade), 6.90 g sodium dodecyl sulfate (SDS, Chemsolute, ≥98.0%) and 10.6 g hydroxyethyl cellulose (HEC, Sigma-Aldrich) in 1.50 L water was prepared in a 2 L round flask via mild stirring at 50 °C, until the HEC was dissolved. After allowing to cool down to room temperature, a suspension of 160 g·L$^{-1}$, oleic acid-stabilized $Fe_3O_4$ nanoparticles (11.5 wt% organic content) in 50 mL cyclohexane (Chemsolute, ≥99.8%) was slowly added to the aqueous solution with a syringe while strongly agitating at 13500 rpm using an Ultra-Turrax™ (IKA, Germany) for 1 min. The flask was kept open to allow evaporation of the solvent and the stirring was continued at 60 rpm using a 68 × 24 × 3 mm$^3$ PTFE stirrer shaft attached to a mechanical stirrer. The clusters were separated, after 20 h, from the remaining free NPs in the surfactant solution by magnetically decanting the supernatant, and redispersion in fresh water. Surfactants were removed by washing the supraparticles 6 times with warm 15% ethanol followed by 6 times warm 96% ethanol, until the particles agglomerate in polar solvents like water. The cleaned supraparticles were redispersed in ethyl acetate and dispersed on a silicon wafer via spin coating.

The heat treatment to induce the crosslinking of the organic ligands was conducted in a tube furnace, with a hold temperature of 325 ºC, hold time of 18 min, heating ramp of 1 ºC·min$^{-1}$, under nitrogen ($N_2$) atmosphere.

**SEM and FIB**

Micrographs of the bulk SCNCs fracture surfaces were obtained via scanning electron microscopy (SEM, Zeiss SUPRA 55-VP, Zeiss, Germany), using 1-5 kV acceleration voltage, working distance of 4-7 mm, and through-the-lens detector (TLD) or Everhart–Thornley detector (ETD). The micropillars from the bulk samples were fabricated via focused ion beam (FIB) milling with a gallium ion source (FEI Helios NanoLab G3, Thermo Fisher Scientific, Oregon, USA). For the milling process, currents from 21 nA down to 1.1 pA for rough cuts and subsequent polishing were used. These parameters are selected based on previous studies on analogous materials systems,[15] which allow us to consider the FIB-induced damage and potential degradation of the organic ligands (in their ultra-confined and often crosslinked state) to be negligible for the purposes of this study. Three of each sample type, micropillars from bulk samples and supraparticles, were transferred onto small pins providing sufficient elevation to avoid shadowing during the X-ray analysis. The pins were previously tested to ensure sufficient stability to the heat treatment temperature (325 ºC), since one supraparticle and one pillar were tested via X-rays both before and after heat treatment. No significant alteration of the pins was detected due to the heat treatment. The sample transfer onto the pins and attachment was done in the FIB instrument using a micromanipulator (see, e.g., Fig. S1) and ion beam induced deposition (IBID) of a Pt precursor material via a gas injection system (GIS) respectively.[92] Micrographs of the prepared micropillars and supraparticles were obtained using a scanning electron microscope under high-vacuum mode, with a 5 kV acceleration voltage, 50 pA beam current, a working distance of 7 mm and with a Everhart–Thornley detector (ETD).

**X-Ray Scattering Experiment**

The X-ray scattering experiment was performed at the Coherence Applications beamline P10 at the PETRA III synchrotron source (DESY, Hamburg, Germany). Monochromatic X-rays of 10 keV were focused down to ~2.5 × 1.9 µm$^2$ (horizontal × vertical) at the sample position completely covering the SCNC samples. The sample pin was fixed on a rotation stage, rotating around the vertical axis. At each angular position, the 2D far-field diffraction patterns were recorded by an EIGER X 4M detector positioned 5.0 m downstream from the sample. The sample was rotated by the increment of 1/3° over the range of 180° and, by that, the full 3D diffraction pattern was measured. At each angular position, 5 frames with 1 s exposure each were collected and then averaged. The sample was cooled down using a liquid nitrogen jet to avoid radiation damage of the organic ligands stabilizing the NPs, which could induce the NPs coalescence and destroy the superlattice ordering.

**AXCCA**

Details on the application of the AXCCA technique to a 3D scattered intensity distribution can be found in Ref. [59].[62] The AXCCA is based on the calculation of the cross-correlation function (CCF), defined as

$$C(q_1, q_2, \Delta) = \langle \tilde{I}(q_1)\tilde{I}(q_2)\delta\left(\frac{q_1 \cdot q_2}{|q_1||q_2|} - \cos\Delta\right)\rangle, \qquad (1)$$

where $\tilde{I}(q_1)$ and $\tilde{I}(q_2)$ are the normalized scattered intensities taken at the momentum transfer vectors **q₁** and **q₂**, respectively, with Δ being the relative angle between them. Eq. (1) is averaged over all angular positions of vectors **q₁** and **q₂** with the momentum transfer values $q_1$ and $q_2$, respectively. The intensity $\tilde{I}(q)$ is normalized by the average intensity at the corresponding *q*-value:

$$\tilde{I}(q) = \frac{I(q) - \langle I(q)\rangle_{|q|=q}}{\langle I(q)\rangle_{|q|=q}}. \qquad (2)$$

When analysing (super)crystalline materials, CCFs contain peaks at the relative angles between the Bragg peaks, *i.e.* the angles between a certain pair of the families of (super)crystallographic planes corresponding to the reflections at $q_1$ and $q_2$. These angles provide additional information on the (super)crystalline structure compared to the conventional analysis of the azimuthal intensity profiles.

Given a model of the unit cell with the lattice basis vectors **a₁**, **a₂** and **a₃**, one can calculate the reciprocal basis vectors **b₁**, **b₂** and **b₃**, and thus any reciprocal lattice vector **g** = *h*·**b₂**+*k*·**b₂**+*l*·**b₃**. Then, the angles between any (super)crystallographic planes can be calculated using the dot product of the corresponding reciprocal vectors $cos[\Delta_{ij}] = \frac{g_i \cdot g_j}{|g_i||g_j|}$. The angle gives the expected peak position in the CCF *C*(*q₁*,*q₂*,Δ) calculated for the scattered intensities at the momentum transfer values *q₁* = |**g**ᵢ| and *q₂* = |**g**ⱼ|. By taking into account the (super)lattice symmetry, one can calculate all expected positions of the correlation peaks and optimize the unit cell parameters to fit the experimentally obtained peaks. For the SCNC samples considered in this study, CCFs were calculated between the intensities taken at *q₁*, corresponding to the first bright peak (the exact momentum transfer values *q₁* can differ for different samples), and at $q_2$, varying in the range from 0.4 to 0.9 nm$^{-1}$, with a step size of 0.005 nm$^{-1}$. The resulting CCF maps are then shown in ($q_2$,Δ)-coordinates.

### In-situ heating STEM

The procedure of temperature ramping and holding for in-situ STEM heating was performed in a FEI Talos F200x (Thermo Fisher Scientific, USA) using an in-situ heating holder. The heating rate was 5 ºC·min$^{-1}$ and the holding time 30 min. The selected holding temperatures are 200, 250, 300, 325, 350, 375, 400 ºC, where high-angle annular dark-field (HAADF) micrographs were taken at the beginning, middle and end of each holding period, to minimize the effect of electron beam damage while monitoring the nanostructure evolution. The tilt angle of the lamella varied from -2.7º (before heating) to 11.8º (at 400 ºC), to compensate for its bending with varying temperature, without causing significant contrast change.

### Mechanical tests

After the X-ray measurements, the microsamples were transferred to a fresh Si substrate for the mechanical tests. They were fixed on the Si substrate with Pt deposition via FIB (FEI Helios G3 UC SEM/FIB, Oregon, USA). The microcompression tests were carried out with a flat diamond punch (Synton-MDP, Nidau, CH) in an Agilent Nano Indenter G200 (Agilent, Santa Clara, CA, USA). The samples were tested with loading-unloading cycles with increasing maximum load. Based on previous microcompression tests,[5] the maximum loads of the different cycles were selected to be 0.15, 0.2, 0.25, 0.3 mN, with a final loading step which proceeded until the fracture of the samples. The loading rate was 10$^{-3}$ mN·s$^{-1}$. Three pillars and one supraparticle were tested (all after heat treatment and X-ray analysis). The fracture loads were 0.56, 0.67 and 0.18 mN for the pillars (with the last pillar loaded at the higher rate of 10$^{-2}$ mN·s$^{-1}$), and 0.32 mN for the supraparticle, corresponding to 502, 722, 250 and 300 MPa, respectively. Note that for the supraparticle this is a representative equivalent compressive strength, calculated as the applied load divided by the sphere's equatorial cross-section, since a sphere under this loading condition experiences a distribution of tensile and compressive stresses at different domain areas. A more detailed analysis of the overall mechanical behaviour of nanocomposite SPs has been conducted elsewhere.[93]

**Molecular mechanics simulations**

The magnetite nanoparticle (NP), 4 nm in diameter, was generated via NanoCrystal[94] with the force field parameters taken from our previous work.[95] Undercoordinated Fe ions were hydroxylated. To ensure charge neutrality of the bare magnetite NP, $Fe^{2+}$ and $Fe^{3+}$ ions were randomly distributed employing the charge neutrality equation introduced in our previous work. $Fe^{2+}$ and $Fe^{3+}$ oxidation states were then equilibrated using the oxidation state swap method.[95]

Subsequently, the NPs were functionalized with oleic acid molecules, with each functionalized magnetite NP containing approximately 17,000 atoms. An in-depth study detailing on the application of this approach to the adsorption of organic molecules on magnetite surface has previously been performed.[96] This approach ensures that the oxidation states adapt to the surrounding electrostatic environment, while maintaining compatibility with common biomolecular force fields. The hydroxylated nanoparticle was equilibrated using a hybrid Monte Carlo/Molecular Dynamics (MC/MD) method.[95] The NP was subsequently functionalized with OLEC molecules.[97] OLEC was described using GAFF[98] parameters and RESP[99] charges. The resulting functionalized NP was further used as a building block for the SCNCs, to build and equilibrate an *fcc* SCNC. The *fcc* and *hcp* crystal structure have as stacking order "ABC" and "AB" sequences, respectively. From the equilibrated *fcc* SCNC, a "ABCABCABABAB" system was created. All simulations were performed with LAMMPS.[100] More details on the specific simulations protocols and force field details are provided in the SI section 5. Beside an optical inspection of the simulated structures, Steinhardt bond order parameters[79] were used to determine the local supercrystalline structure. These parameters take into account the local orientational order using spherical harmonics. Here, they are calculated employing the "compute orient-order/atom command" of LAMMPS for the nearest-neighbour shell of the NPs using their centre of mass for the analysis. Particularly, the $q_4$ and $q_6$ Steinhardt bond order parameters were employed in this study, since they are highly effective for distinguishing between *fcc* and *hcp* local structures. For perfect lattices, fcc corresponds to $q_4$ = 0.19 and $q_6$ = 0.575, while hcp corresponds to $q_4$ = 0.097 and $q_6$ = 0.484. For non-ideal systems the values often deviate from the perfect ones and are found in a broader range, particularly for hcp $q_4$ is often between 0.05 and 0.15, while the others remain closer to the ideal values.[101,102]

**Acknowledgements**

The authors gratefully acknowledge the financial support from the Deutsche Forschungsgemeinschaft (DFG, German Research Foundation), project numbers GI 1471/1-1 and 192346071-SFB 986. D.G. thanks the support from the Institute of Complex Molecular Systems (ICMS) at TU/e. The use of the FIB dual beam instrument at the DESY NanoLab granted by BMBF under grant no. 5K13WC3 (PT-DESY) is acknowledged. We are thankful for Dr. Junwei Wang and Prof. Nicolas Vogel for their valuable support with the analysis of the optimized particle positions of SP anti-Mackay structures. We acknowledge DESY (Hamburg, Germany), a member of the Helmholtz Association HGF, for the provision of experimental facilities. Parts of this research were carried out at PETRA III, beamline P10. Beamtime was allocated for Proposal I-2019118.

**Author contributions**

D.L. devised and performed the X-ray scattering study and AXCCA analysis, C.Y. devised and performed the in-situ STEM study, E.G. devised and performed the MD study, G.B.V.-F. and R.H.M. supervised the MD study, H.S. supported the analysis of the in-situ STEM study and the manuscript writing, A.P. prepared the SP samples and supported the X-ray scattering study, B.B. prepared the bulk samples and performed the mechanical testing, Y.Y.K., D.A., F.W., M.S., supported the X-ray scattering study, T.K., S.S.R., S.K., M.R. performed and supervised EM and FIB studies, T.F.K. supported and supervised the X-ray scattering study and the EM study, G.A.S. supported the design of the study, A.S. and I.A.V. devised and

supervised the study,  D.G. devised, supervised and coordinated the study, D.L. and D.G. wrote the article. All authors reviewed the article.

**Declaration of competing interests**

The authors declare no competing interests.

**Data availability**

Data available on request from the authors.

# Supplementary Information for

## Defect migration in supercrystalline nanocomposites

Dmitry Lapkin[1,*,i], Cong Yan[2], Emre Gürsoy[3], Hadas Sternlicht[4], Alexander Plunkett[5], Büsra Bor[5], Young Yong Kim[1], Dameli Assalauova[1,iii], Fabian Westermeier[1], Michael Sprung[1], Tobias Krekeler[6], Surya Snata Rout[6,ii], Martin Ritter[6], Satishkumar Kulkarni[7], Thomas F. Keller[7,8], Gerold A. Schneider[5], Gregor B. Vonbun-Feldbauer[3,9], Robert H. Meissner[3,9], Andreas Stierle[7,8], Ivan A. Vartanyants[1], Diletta Giuntini[2,5,*]

[1] Photon Science, Deutsches Elektronen-Synchrotron (DESY), Hamburg, Germany

[2] Department of Mechanical Engineering, Eindhoven University of Technology, Eindhoven, Netherlands

[3] Institute of Interface Physics and Engineering, Hamburg University of Technology, Hamburg, Germany

[4] Department of Materials Science and Engineering, Pennsylvania State University, University Park, PA, USA

[5] Institute of Advanced Ceramics, Hamburg University of Technology, Hamburg, Germany

[6] Electron Microscopy Unit, Hamburg University of Technology, Hamburg, Germany

[7] Centre for X-ray and Nano Science, Deutsches Elektronen-Synchrotron (DESY), Hamburg, Germany

[8] Department of Physics, University of Hamburg, Hamburg, Germany

[9] Institute of Surface Science, Helmholtz-Zentrum-Hereon, Geesthacht, Germany

[*] Corresponding authors

[i] Current address: Institute of Applied Physics, University of Tübingen, Auf der Morgenstelle 10, 72076 Tübingen, Germany

[ii] Current address: School of Earth & Planetary Sciences, National Institute of Science Education and Research, Jatani, 752050 Khurda, India; Homi Bhabha National Institute, Training School Complex, Anushaktinagar, 400094 Mumbai, India

[iii] Current address: Constructor University, Campus Ring 1, D-28759 Bremen, Germany

1. **Micropillars and supraparticles prepared for X-ray analysis**

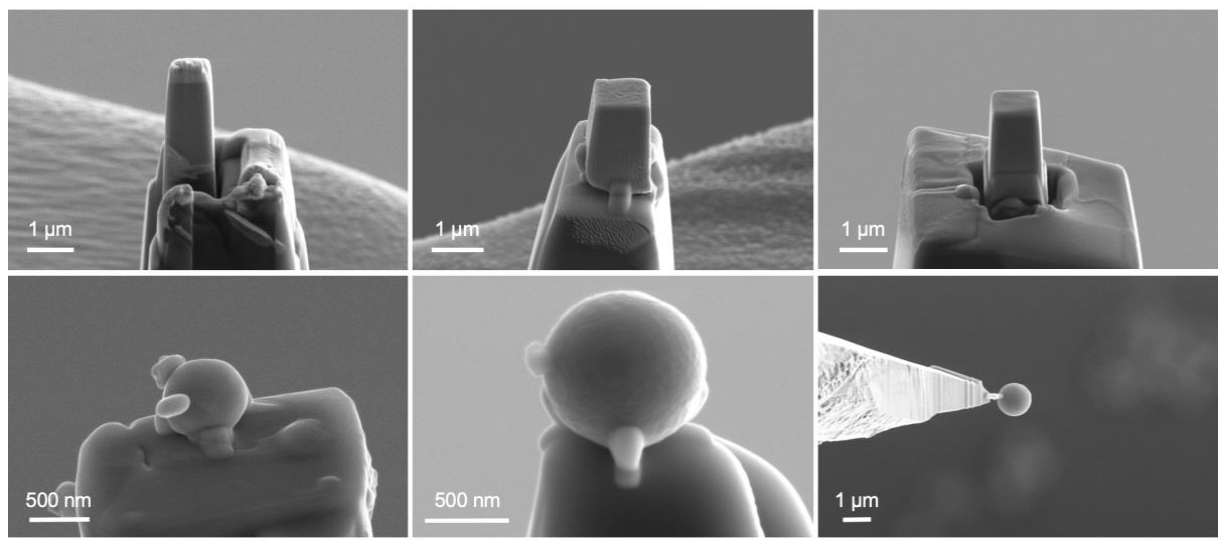

**Fig. S1:** Secondary electron images of micropillars ("Pillars") from bulk samples mounted on top of a pin suitable for the 3D X-ray analysis (top row) and supraparticles ("SPs") during and after the transfer via the micromanipulator onto the pin inside the dual beam focused ion beam instrument.

## 2. Microcompression tests

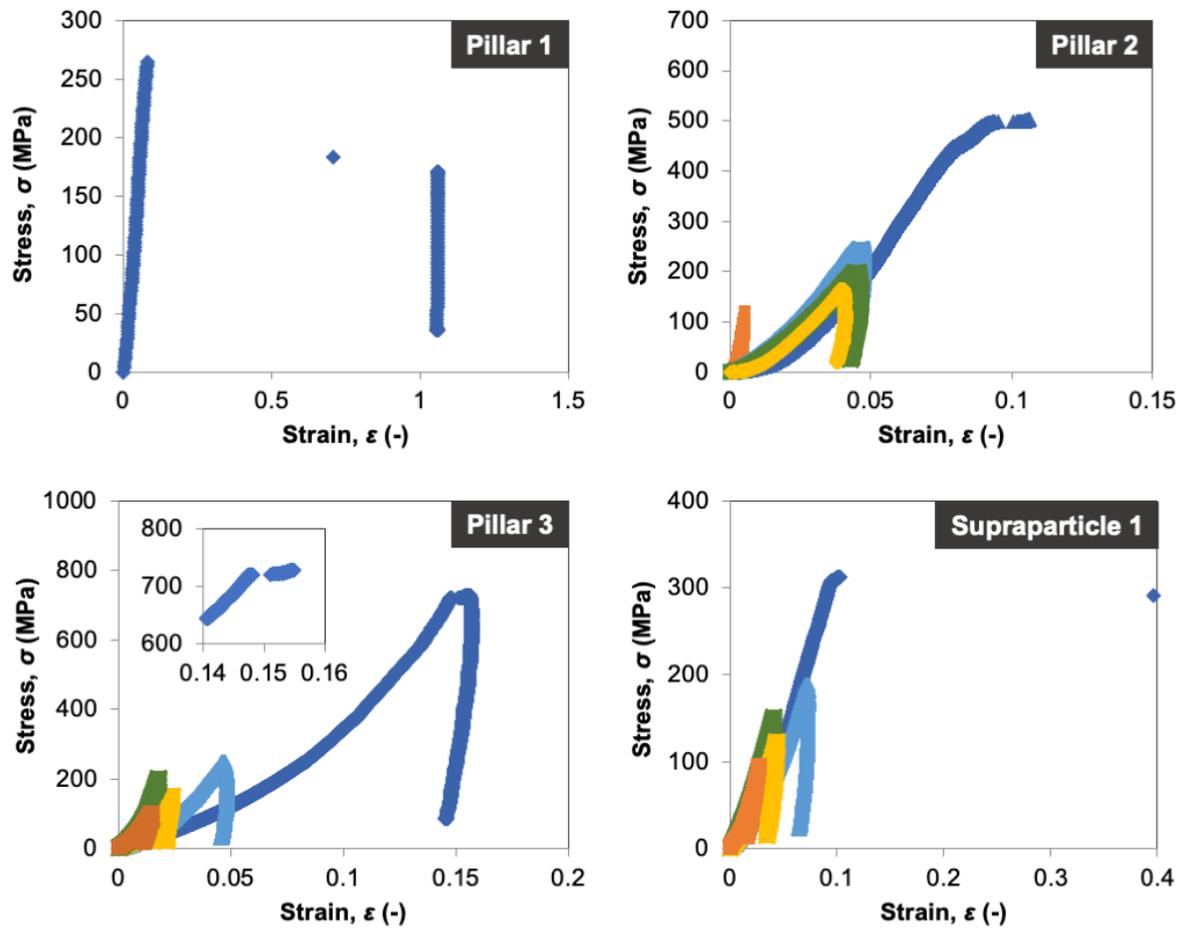

**Fig. S2:** Microcompression stress-strain curves for Pillars and SP, all in heat-treated (crosslinked) state. The other two SPs were damaged during sample transfer after the AXCCA analysis and could thus not be tested. The tests were conducted by loading and unloading each sample in compression in steps, until fracture. The colours indicate different test cycles, the dark blue colour indicates the cycle of final fracture.

## 3. AXCCA optimization of the unit cell parameters of Pillar and SP samples

The unit cell parameters were optimized as described in the Methods section in the main text.[1] Given the primitive unit cell parameters $a'$, $b'$, $c'$ and $\alpha'$, $\beta'$, $\gamma'$, we calculated the real and reciprocal basis vectors $\mathbf{a_1}$, $\mathbf{a_2}$, $\mathbf{a_3}$ and $\mathbf{b_1}$, $\mathbf{b_2}$, $\mathbf{b_3}$. Then, we calculated all expected peak positions $(q_2,\Delta)$ in the CCFs $C(q_1,q_2,\Delta)$ for the given unit cell parameters. The mean value $\langle C \rangle$ of the CCFs $C(q_1,q_2,\Delta)$ at the expected peak positions $(q_2,\Delta)$ was used as a metric. We found a global maximum of the metric and then tuned the unit cell parameters in a small range around the global maximum. The dependence of the $\langle C \rangle$ value on the parameters is shown in, e.g., Fig. S3. The positions of the maximum are treated as the average unit cell parameters, the HWHM (half width at half maximum) are treated as the distribution width and given in Tables 1 and 2 in the main text. One should note that the width of the peak is defined not only by the microstrain (the unit cell parameters distribution in the sample), but also by the sample size. Therefore, the obtained values give the upper estimations for the parameter distribution widths.

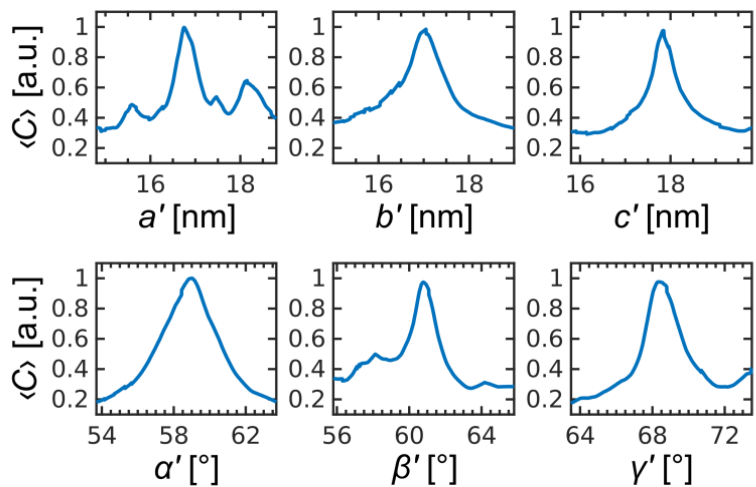

**Fig. S3. Angular X-ray Cross-Correlation Analysis (AXCCA) of Pillar 1.** Mean correlation values at the peak positions in the CCF map for Pillar 1 (shown in Fig. 2e in the main text) for a triclinic structure with the unit parameters $a$, $b$, $c$, $\alpha$, $\beta$, $\gamma$. Each parameter is tuned separately, when all other are set to the ones giving the maximum value of the correlation.

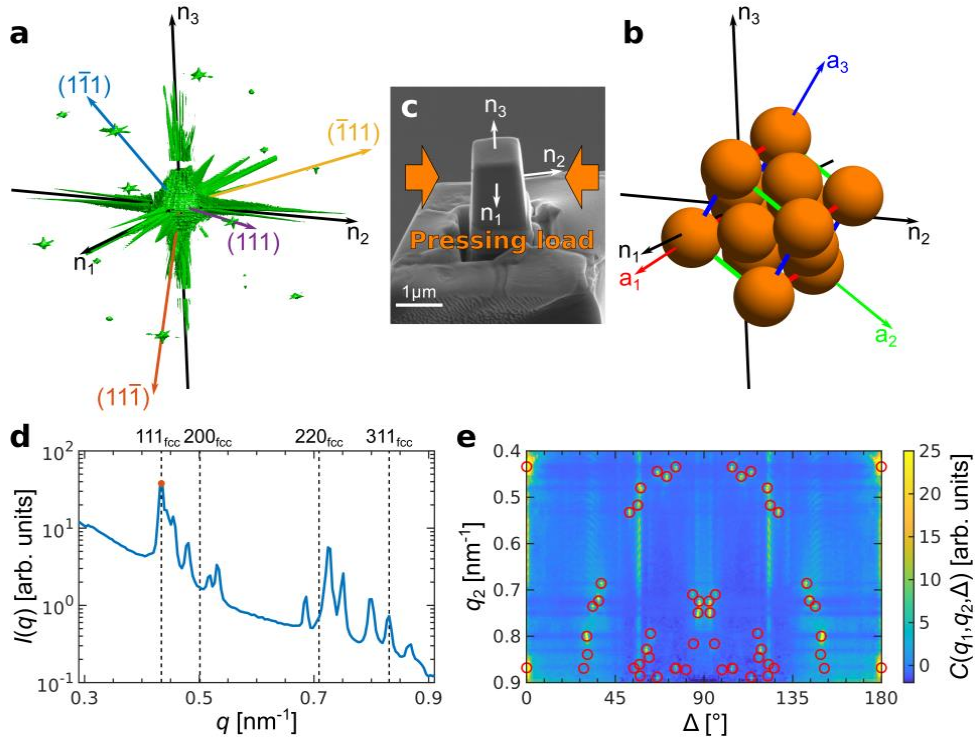

**Fig. S4. Angular X-ray Cross-Correlation Analysis (AXCCA) of Pillar 2. a** An isosurface of the measured scattered intensity distribution in 3D reciprocal space for Pillar 2. Four $(111)_{fcc}$ directions of the reciprocal lattice of a distorted *fcc* lattice are indicated by coloured arrows as well as the normal vectors $\mathbf{n_1}$, $\mathbf{n_2}$ and $\mathbf{n_3}$ to the pillar walls deduced from the intensity "spikes" orientation. **b** Orientation of a distorted *fcc* unit cell with respect to the pillar walls in real space. **c** An SEM image of the same pillar with indicated directions $\mathbf{n_1}$, $\mathbf{n_2}$ and $\mathbf{n_3}$. The uniaxial stress direction applied during the sample preparation is also indicated. **d** Azimuthally averaged intensity profile of the 3D scattered intensity of Pillar 2, with the red point indicating $q_1 = 0.433$ nm$^{-1}$, used for the calculation of the cross-correlation functions (CCFs). The peak positions of an ideal *fcc* structure are indicated with vertical dashed lines **e** CCFs $C(q_1,q_2,\Delta)$, calculated for $q_1$ (indicated in **d**) and $q_2$ in the range of 0.4 – 0.9 nm$^{-1}$, stacked along the vertical axis $q_2$, with the peak positions for the optimized unit cell parameters marked by red circles.

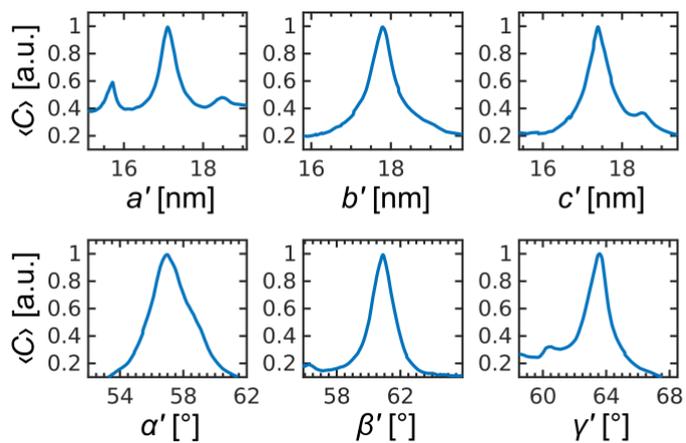

**Fig. S5. Angular X-ray Cross-Correlation Analysis (AXCCA) of Pillar 2.** Mean correlation values at the peak positions in the CCF map for Pillar 2 (shown in Fig. S4e) for a triclinic structure with the unit parameters *a*, *b*, *c*, *α*, *β*, *γ*. Each parameter is tuned separately, when all other are set to the ones giving the maximum value of the correlation.

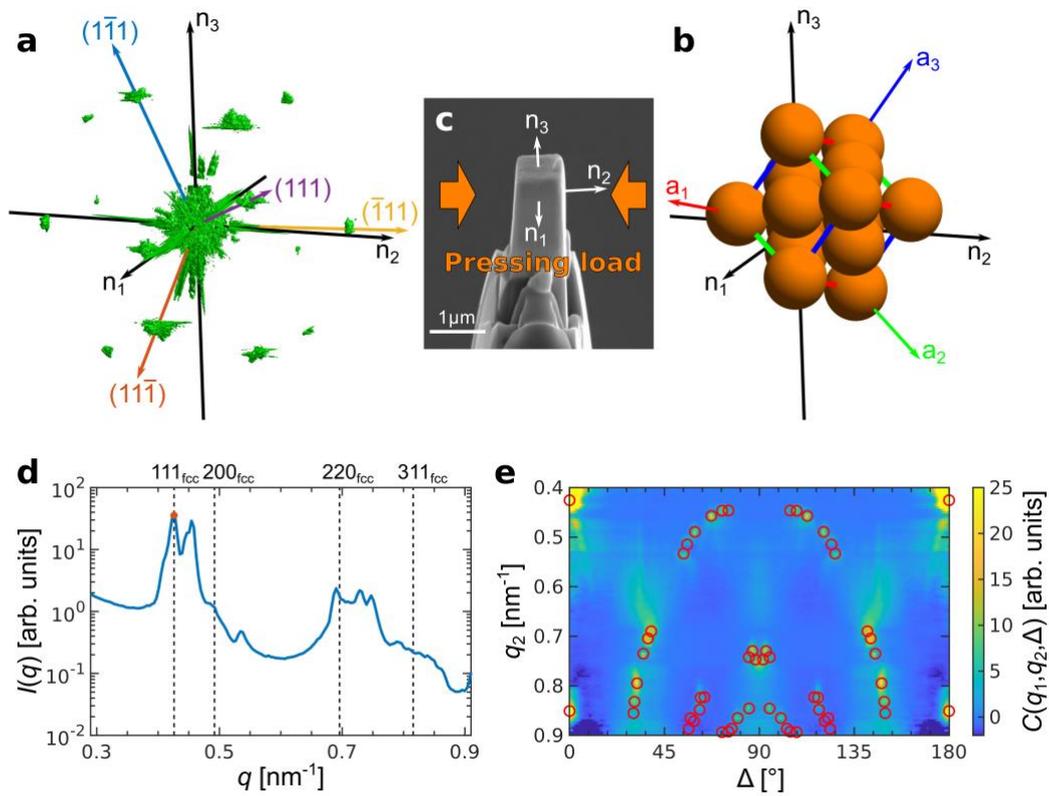

**Fig. S6. Angular X-ray Cross-Correlation Analysis (AXCCA) of Pillar 3 before HT. a** An isosurface of the measured scattered intensity distribution in 3D reciprocal space for Pillar 3 before HT. Four (111)$_{fcc}$ directions of the reciprocal lattice of a distorted *fcc* lattice are indicated by coloured arrows as well as the normal vectors $n_1$, $n_2$ and $n_3$ to the pillar walls deduced from the intensity "spikes" orientation. **b** Orientation of a distorted *fcc* unit cell with respect to the pillar walls in real space. **c** An SEM image of the same pillar with indicated directions $n_1$, $n_2$ and $n_3$. The uniaxial stress direction applied during the sample preparation is also indicated. **d** Azimuthally averaged intensity profile of the 3D scattered intensity of Pillar 3, with the red point indicating $q_1 = 0.426$ nm$^{-1}$, used for the calculation of the cross-correlation functions (CCFs). Peak positions for an ideal *fcc* structure are indicated with vertical dashed lines. **e** CCFs $C(q_1,q_2,\Delta)$, calculated for $q_1$ (indicated in **d**) and $q_2$ in the range of 0.4 – 0.9 nm$^{-1}$, stacked along the vertical axis $q_2$, with the peak positions for the optimized unit cell parameters marked by red circles.

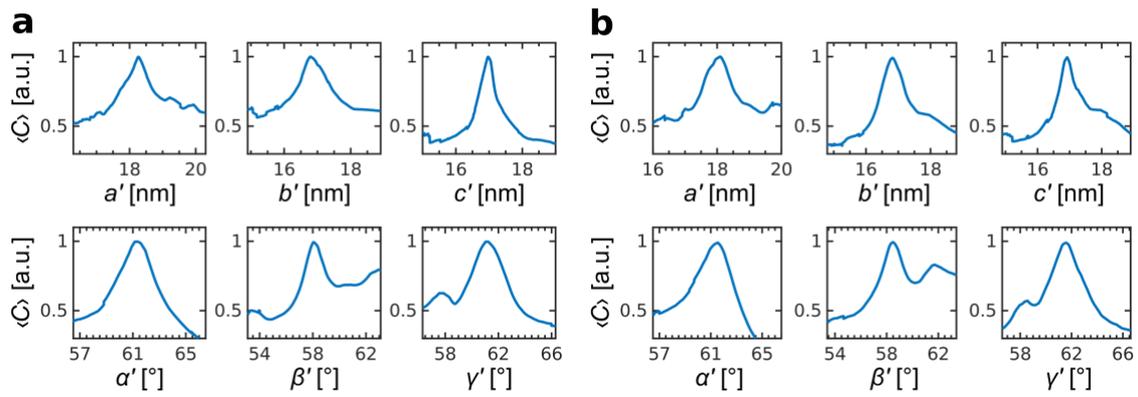

**Fig. S7. Angular X-ray Cross-Correlation Analysis (AXCCA) of Pillar 3 before (a) and after (b) heat treatment.** Mean correlation values at the peak positions in the CCF map for Pillar 3 (shown in Fig. S6e) for a triclinic structure with the unit parameters *a*, *b*, *c*, *α*, *β*, *γ*. Each parameter is tuned separately, when all other are set to the ones giving the maximum value of the correlation.

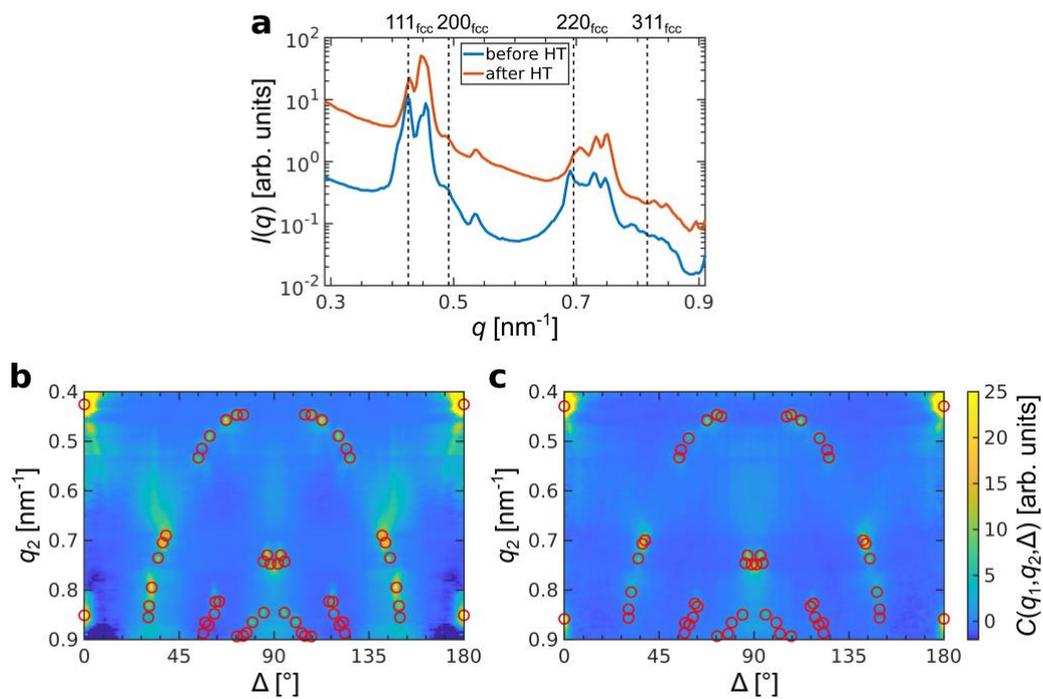

**Fig. S8: Angular X-ray Cross-Correlation Analysis (AXCCA) of Pillar 3 before and after heat treatment. a** Azimuthally averaged intensity profiles of the 3D scattered intensities of Pillar 3 before (blue line) and after (red line) heat treatment. The profiles are shifted vertically for clarity. Peak positions of an ideal *fcc* structure are indicated with vertical dashed lines. **b, c** CCFs $C(q_1,q_2,\Delta)$ calculated for $q_1$ corresponding to the $111_{fcc}$ Bragg peak and $q_2$ in the range of 0.4 – 0.9 nm$^{-1}$ before (**b**) and after (**c**) heat treatment. The CCFs are shown as a heat map stacked along vertical axis $q_2$.

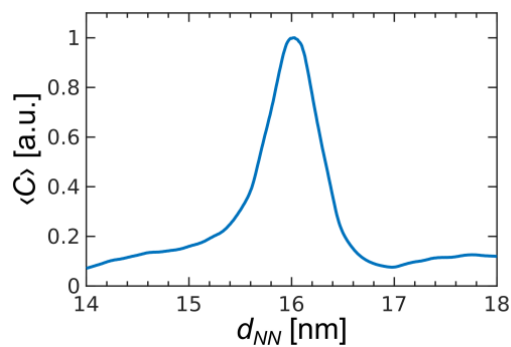

**Fig. S9. Angular X-ray Cross-Correlation Analysis (AXCCA) of SP 1.** Mean correlation values at the peak and arc positions in the CCF map for SP 1 (shown in Fig. 3c in the main text) for an *fcc* structure as a function of the nearest neighbor distance $d_{NN}$.

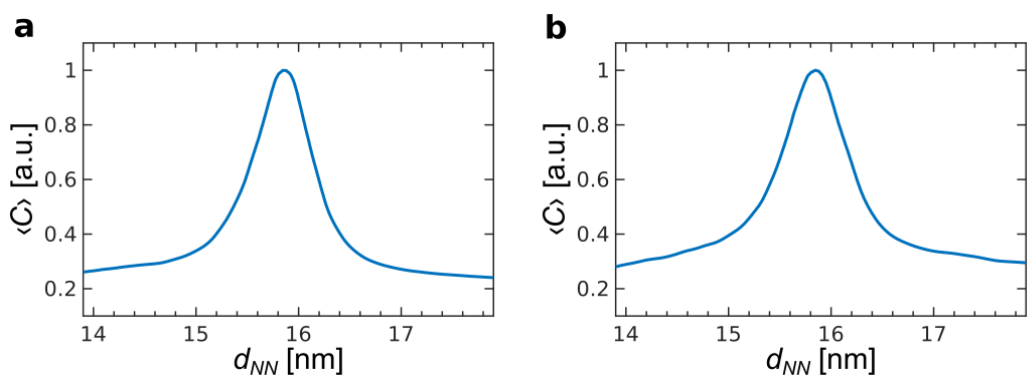

**Fig. S10. Angular X-ray Cross-Correlation Analysis (AXCCA) of SP 3 before (a) and after (b) heat treatment.** Mean correlation values at the peak and arc positions in the CCF map for SP 3 (shown in Figs. 4d,e in the main text) for an *fcc* structure as a function of the nearest neighbor distance $d_{NN}$.

## 4. AXCCA of the anti-Mackay supraparticle

The optimized particle positions of the anti-Mackay structure in a SP of the corresponding size were obtained based on previous works from FAU Erlangen.[2–4] The particle positions were used to calculate 2D X-ray scattering patterns for the same details as in the experiment (see the Methods section of the main text). The 2D pattern were simulated using the MOLTRANS software. The obtained 2D patterns were interpolated onto a 3D grid to obtain the entire 3D scattered intensity distribution. The CCFs $C(q_1,q_2,\Delta)$ were then calculated for $q_1$ corresponding to the $111_{fcc}$ Bragg peak and $q_2$ varied in the range of 0.4 – 0.9 nm$^{-1}$. The resulting CCFs of the simulated intensity distribution is shown in Fig. S11b for comparison with the experimental CCFs shown in Fig. S11a. The regions of the $111_{fcc}$ and $200_{fcc}$ Bragg peaks ($q \approx 0.475$ nm$^{-1}$ and $q \approx 0.775$ nm$^{-1}$, respectively) are very similar and contain the same characteristic peaks for the anti-Mackay structure, which are distinctive from those of a simple $fcc$ structure. On the other hand, the CCFs calculated for the simulated intensity distribution contain many additional peaks in the $q$-range between the Bragg peaks ($q = 0.5 – 0.75$ nm$^{-1}$) because the simulation did not include any noise hiding the correlations in the experimental data.

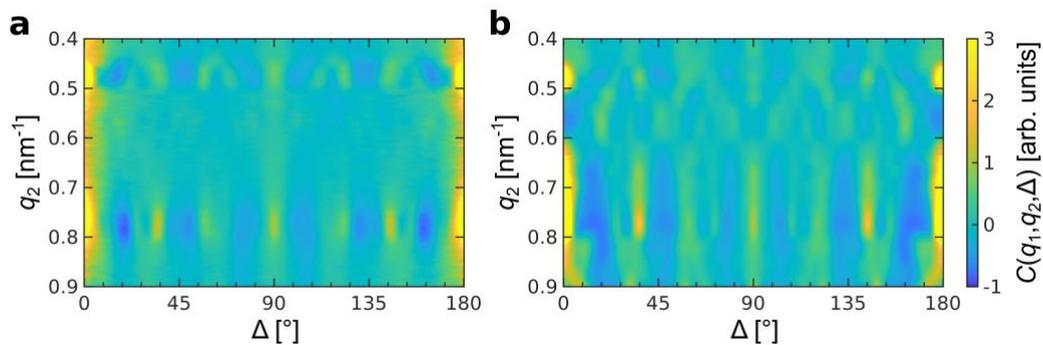

**Fig. S11. Angular X-ray Cross-Correlation Analysis (AXCCA) of SP 2 with anti-Mackay structure.** CCFs $C(q_1,q_2,\Delta)$ calculated for $q_1$ corresponding to the $111_{fcc}$ Bragg peak and $q_2$ in the range of 0.4 – 0.9 nm$^{-1}$ for experimental (**a**) and simulated (**b**) 3D scattered intensity distributions. The CCFs are shown as a heat map stacked along the vertical axis $q_2$.

## 5. All-atom simulations

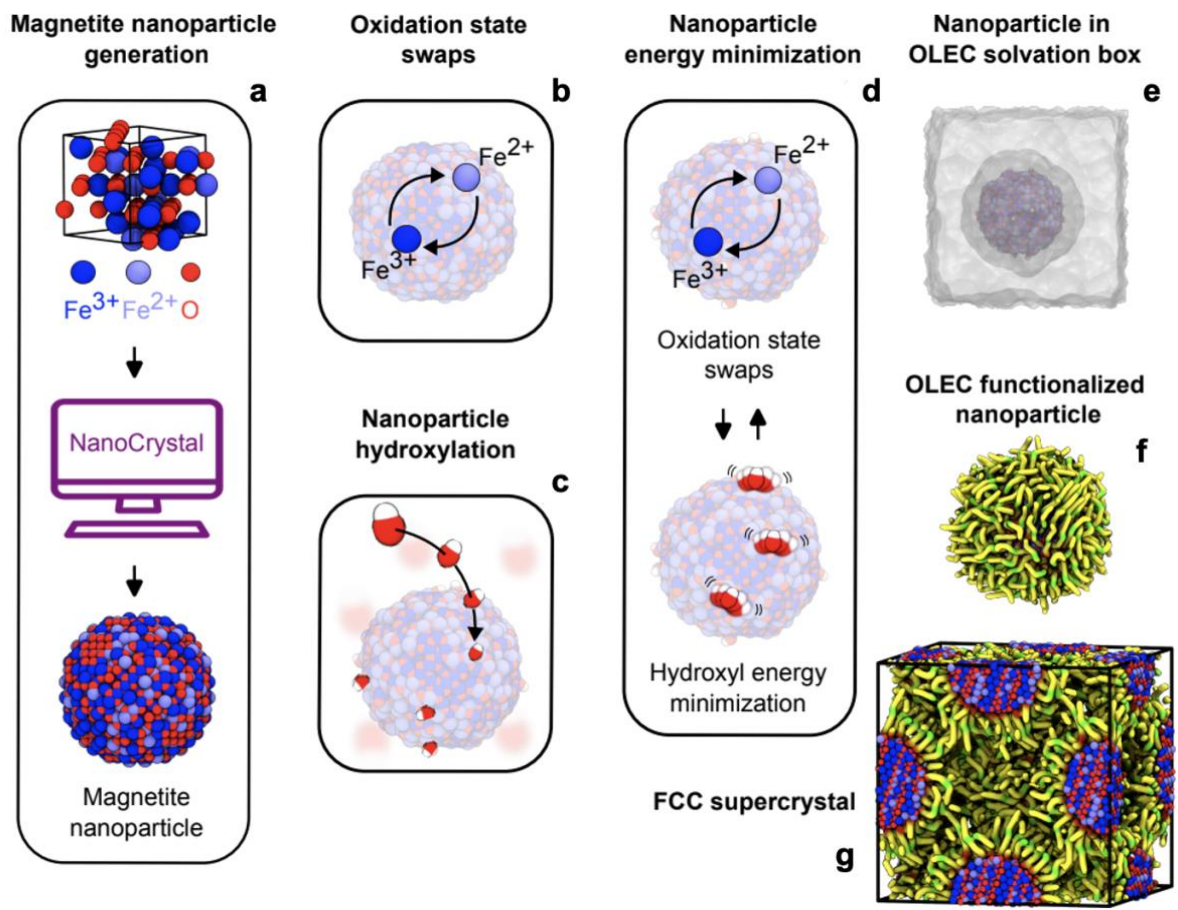

**Fig. S12. Multiscale workflow for generating all-atom OLEC functionalized magnetite SCNCs. a** Generation of a magnetite NP, **b** oxidation state swaps and **c** hydroxylation of NP; **d** Minimization of potential energy; **e** Functionalization of NP in OLEC solvation box; **f** OLEC functionalized NP building block to construct the SCNC (**g**).

**Magnetite nanoparticle generation**

A spherical magnetite nanoparticle (NP) with a diameter of 4 nm based on an $Fd\bar{3}m$ unit cell (see Fig. S12a) was generated using NanoCrystal[5]. This is a web-based tool that generates NP models using their crystal structure, desired size, preferred growth planes and energies. The coordination polyhedron option was not used. After NP generation, the partial atomic charges of Fe ions were randomly distributed to ensure charge neutrality using the charge neutrality equation for non-stoichiometric magnetite particles or surfaces.[6] Forcefield parameters describing magnetite NP were taken from previous work.[7]

**NP oxidation state equilibration**

Oxidation states of magnetite NPs were distributed using oxidation state swaps[6] (see Fig. S12b). The oxidation state swap method is an atomistic simulation method that was developed specifically for modelling magnetite structures while ensuring compatibility to common biomolecular force fields. It is based on exchanging oxidation states of Fe ions using a Monte Carlo (MC) approach, with or without the combination of Molecular Dynamics (MD). To minimize the oxidation states of the NPs inside a 150 × 150 × 150 Å$^3$ simulation box with non-periodic boundary conditions MC swaps were performed using LAMMPS[8]. For pairwise interactions, instead of using a long-range solver, a relatively large cut-off of 45 Å was used. This cut-off ensured that the interactions between magnetite ions on opposite sides of the NP were taken into account. $n_{swaps} \approx (n_{Fe})^2$ swaps were performed at $T^{MC}$ = 300 K temperature.

**NP hydroxylation and energy minimization**

With the setting "coordination polyhedral" turned off, the spherical NP produced using NanoCrystal[5] is not stoichiometric. Particularly, the generated NP contained eleven oxygen atoms less than for the ideal magnetite stoichiometry. To ensure magnetite stoichiometry, the eleven least coordinated Fe ions on the surface were hydroxylated (Fig. S12c). Energy minimization and oxidation state swaps were performed on the hydroxylated NP in cyclic order. During the energy minimization step, magnetite ions were kept frozen. After each minimization step, oxidation state swaps were applied to allow the Fe ions to adapt to the updated hydroxyl geometry (Fig. S12d). At each oxidation state swap step, approximately $n_{swaps} \approx (n_{Fe})^2$ were performed at $T^{MC}$ = 300 K. In total, ten cycles of energy minimization and oxidation state swaps were performed. For pairwise interactions non-periodic boundary conditions were used with a relatively large cut-off of 55 Å to ensure that interactions between hydroxides on opposite sides of the NP were accounted for. Following potential energy minimization, the hydroxylated NP was equilibrated using hybrid Monte Carlo/Molecular Dynamics (MC/MD)[6] for 1 ns with a timestep of 0.5 fs. At each MD step, one MC swap was attempted, and both the $T^{MD}$ and $T^{MC}$ were held at 300 K. Pairwise interactions of bonded atoms in hydroxyl groups were scaled to zero.

**Oleic acid solvation box generation and equilibration**

Oleic acid (OLEC) molecules, i.e. C18H34O2, were modelled using GAFF[9] parameters and RESP charges.[10] A 150 × 150 × 150 Å$^3$ simulation box was randomly filled with 2000 OLEC molecules. The OLEC box was equilibrated in the *NpT* ensemble starting from 300 K temperature and 100 atm pressure and gradually lowering the pressure until 1 atm over a period of 1 ns while keeping the temperature constant. Running the simulation at this relatively high pressure avoids potential OLEC aggregation (bubble formation), while allowing OLEC molecules to reorient. After the initial equilibration, a follow-up *NpT* simulation was performed for 1 ns at 300 K and 1 atm pressure to equilibrate the liquid OLEC box. As a result, an OLEC density of 0.88 g·cm$^{-3}$ was obtained which agrees well with the experimental OLEC density at room temperature of 0.89 g·cm$^{-3}$. For pairwise interactions, a PPPM[11] solver with a precision of 10$^{-6}$ and a real-space cutoff of 12 Å was used. For bonded atoms, pairwise interactions associated with 1-2, 1-3 and 1-4 terms were scaled to fit GAFF.

**Magnetite NP functionalization**

A cavity approximately as large as the hydroxylated NP was carved out of the liquid OLEC box. The hydroxylated NP was then placed in this cavity (see Fig. S12e). Initially the combined system was sampled at 500 K temperature for 10 ns using an *NVE* ensemble combined with a CSVR[12] thermostat, allowing OLEC molecules to reorient. Subsequently, the system was sampled via an *NpT* ensemble starting from 500 K and 100 atm pressure which was gradually reduced to a temperature of 300 K over 5 ns while the pressure was kept constant. This was followed by another *NpT* equilibration at 300 K and 100 atm, during which the pressure was gradually lowered to 1 atm over 5 ns while the temperature was kept constant. For pairwise interactions, a PPPM[11] solver with a precision of 10$^{-6}$ and a real-space cutoff of 12 Å and above mentioned 1-2, 1-3, 1-4 scaling factors of GAFF were used. To calculate the OLEC coverage, the number of carboxylic hydrogens within 6 Å of the NP surface was determined and divided by the NP's surface area (see Fig S13). While the first coordination shell of carboxylic hydrogens is within 2 Å from the magnetite surface, the larger distance of 6 Å was specifically chosen to also include OLEC molecules close to surface hydroxide groups.

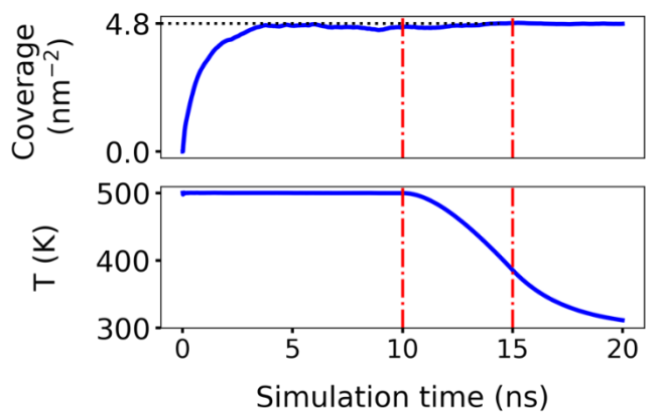

**Fig. S13. Oleic acid coverage on the magnetite nanoparticle (NP).** The upper figure shows the time-dependent evolution of the oleic acid coverage on the magnetite NP, calculated as the number of carboxylic hydrogens within 6 Å of the NP surface divided by NP surface area. The lower figure depicts the temperature during the molecular dynamics simulation.

**SCNC generation and equilibration**

The non-bonded OLEC molecules were removed from the system, resulting in an 16925-atom OLEC-functionalized magnetite NP (Fig. S12f), which was used as the building block to construct SCNCs (Fig. S12g). An *fcc* SCNC was generated using Moltemplate[13] and the [111] direction of the *fcc* structure was aligned to the z-axis of the simulation box corresponding to the close-packed direction. Subsequently, the structure was equilibrated for 10 ns using *NpT* ensemble at 1 atm pressure and 300 K temperature. To use a relatively large timestep of 1 fs, all bonds involving hydrogen were constrained to their original length using the SHAKE algorithm. Contrary to before, in order to speed up the much larger simulations, A PPPM[11] solver with a precision of $10^{-4}$ was used.

## 6. In situ heat treatment in the TEM

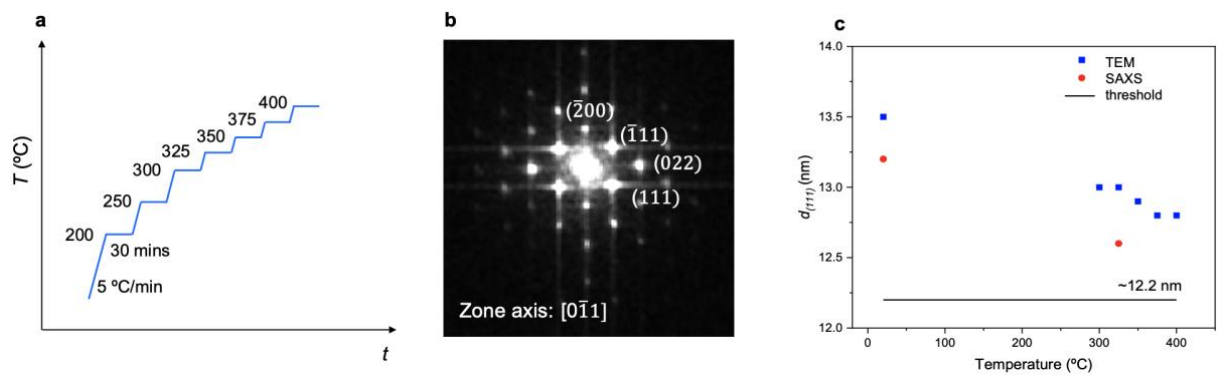

**Fig. S14. In situ heat treatment in the TEM. a** Temperature profile applied during the experiment. **b** Identification of supercrystalline planes reflections and indication of zone axis. **c** Changes in the distance between (111) planes with increasing heat treatment temperature. TEM values are estimated as distances between inverse FFT fringes, SAXS ones are global values obtained for the whole SCNC sample from which the lamella is extracted. The discrepancy between the two is attributed to the tilt and rotation of the lamella during the *in-situ* test, and to the fact that SAXS provides a global average over the whole mm-scale sample. In both cases, the inter-NP distances remain above the 12.2 nm limit that corresponds to the onset of sintering (NPs in contact with each other).